\documentclass[journal=nalefd,manuscript=article,layout=twocolumn,email=true]{achemso}
\usepackage{graphicx}
\usepackage{dcolumn}
\usepackage{bm}
\usepackage{braket}
\usepackage{braket}
\usepackage{amsmath}
\usepackage{amssymb}
\usepackage{lmodern} % for better accents (Czech, etc)
\usepackage{ulem} %for \sout{text}
\usepackage{array} %for making a beautiful table

\newcommand{\onlinecite}[1]{\hspace{-1 ex}\nocite{#1}\citenum{#1}}

\title{Bright-Dark Exciton Interplay Evidenced by Spin Polarization in CdSe/CdMnS Nanoplatelets}

\author{Elena~V.~Shornikova}
\affiliation{Experimentelle Physik 2, Technische Universit\"at Dortmund,~44227 Dortmund, Germany}
\email{elena.shornikova@tu-dortmund.de}
\author{Dmitri~R.~Yakovlev}
\affiliation{Experimentelle Physik 2, Technische Universit\"at Dortmund,~44227 Dortmund, Germany}
\email{dmitri.yakovlev@tu-dortmund.de}
\author{Danil~O.~Tolmachev}
\affiliation{Experimentelle Physik 2, Technische Universit\"at Dortmund,~44227 Dortmund, Germany}
\author{Mikhail A. Prosnikov}
\affiliation{High Field Magnet Laboratory (HFML-EMFL), Radboud University, 6525 ED Nijmegen, The Netherlands}
\author{Peter C. M. Christianen}
\affiliation{High Field Magnet Laboratory (HFML-EMFL), Radboud University, 6525 ED Nijmegen, The Netherlands}
\author{Sushant~Shendre}
\affiliation{LUMINOUS! Center of Excellence for Semiconductor Lighting and Displays, School of Electrical and Electronic Engineering, School of Physical and Materials Sciences, Nanyang Technological University, 639798 Singapore}
\author{Furkan Isik}
\affiliation{LUMINOUS! Center of Excellence for Semiconductor Lighting and Displays, School of Electrical and Electronic Engineering, School of Physical and Materials Sciences, Nanyang Technological University, 639798 Singapore}
\author{Savas~Delikanli}
\affiliation{LUMINOUS! Center of Excellence for Semiconductor Lighting and Displays, School of Electrical and Electronic Engineering, School of Physical and Materials Sciences, Nanyang Technological University, 639798 Singapore}
\alsoaffiliation{Department of Electrical and Electronics Engineering, Department of Physics, UNAM -- Institute of Materials Science and Nanotechnology, Bilkent University, 06800 Ankara, Turkey}
\author{Hilmi~Volkan~Demir}
\affiliation{LUMINOUS! Center of Excellence for Semiconductor Lighting and Displays, School of Electrical and Electronic Engineering, School of Physical and Materials Sciences, Nanyang Technological University, 639798 Singapore}
\alsoaffiliation{Department of Electrical and Electronics Engineering, Department of Physics, UNAM -- Institute of Materials Science and Nanotechnology, Bilkent University, 06800 Ankara, Turkey}
\author{Manfred~Bayer}
\affiliation{Experimentelle Physik 2, Technische Universit\"at Dortmund,~44227 Dortmund, Germany}

\usepackage{color}

\newcommand{\al}[1]{\textcolor{red}{ #1}}

\begin{document}

\begin{abstract}
\includegraphics{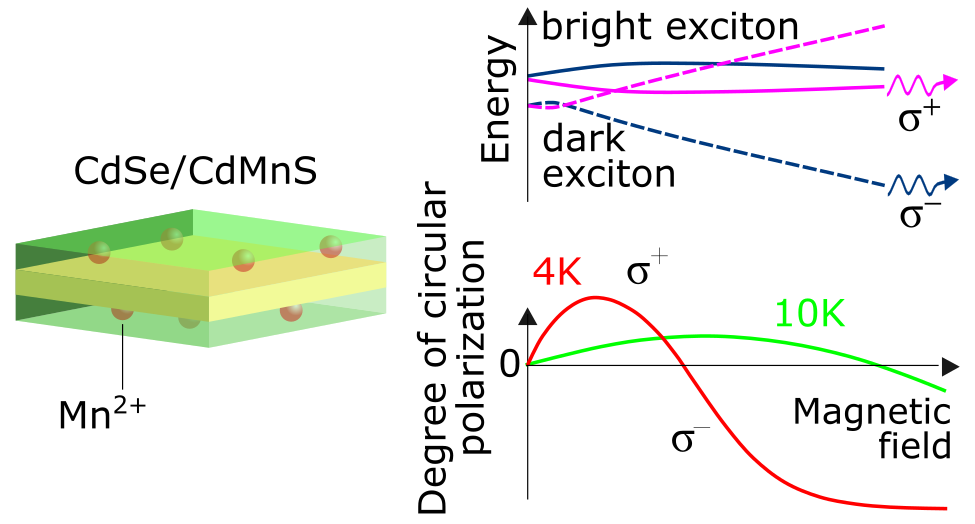}

Diluted magnetic semiconductor (DMS) colloidal nanocrystals demonstrate remarkable magneto-optical properties. However, the behavior of circular polarization of their emission in high magnetic fields remains unclear. We measure magneto-optical properties of colloidal CdSe/CdMnS nanoplatelets in high magnetic fields up to 30 T and at cryogenic temperatures. The degree of circular polarization of photoluminescence demonstrates non-monotonous behavior in a magnetic field. In low magnetic fields, the polarization degree is positive, due to an exchange interaction of excitons with localized spins of magnetic Mn$^{2+}$ ions. The exchange interaction is strong enough to overcome the intrinsic Zeeman splitting, which provides negatively polarized emission in nonmagnetic CdSe/CdS nanoplatelets. After reaching a maximum the polarization degree starts to decrease and reverses the sign to negative in high magnetic fields. The critical magnetic field, in which the sign is reversed, increases when temperature is elevated. We develop a model, which explains this behavior by an interplay of bright and dark exciton recombination. 
\end{abstract}

\maketitle

\textbf{Keywords:} diluted magnetic semiconductors, CdSe nanoplatelets, colloidal nanocrystals, magneto-optics, high magnetic fields.

\vspace{0.5cm}
Diluted magnetic semiconductor (DMS) colloidal nanocrystals (NCs) are formed by doping colloidal NCs with transition metal ions, commonly Mn$^{2+}$ ions are used for II--VI semiconductors.\cite{Beaulac2008adv,Bussian2009,Norris2008} The strong exchange interaction between the host semiconductor $sp-$electrons and holes, and the localized transition metal $d-$electrons leads to bright magneto-optical phenomena. These phenomena were first investigated in bulk DMS materials based on II-VI semiconductors like (Cd,Mn)Te, (Zn,Mn)Se, etc., and later in DMS quantum wells and quantum dots grown by molecular-beam epitaxy~\cite{Furdyna1988,Furdyna1988book,Dietl1994ch,Kossut2010}.

Synthesis and understanding of optical properties of colloidal DMS NCs are still at an early stage, while several important results have been already established. The giant Zeeman splitting has been demonstrated in various experiments.\cite{Hoffman2000,Norris2001,Beaulac2008nl-1,Turyanska2014,Shornikova2020acsn} In contrast to the intrinsic Zeeman splitting, which is observed in nonmagnetic NCs, the exchange interaction with the magnetic impurities provides an additional spin splitting, which can be several orders of magnitude larger than the intrinsic splitting. The exchange interaction of excitons with Mn$^{2+}$ ions has been proven by optically detected magnetic resonance (ODMR)\cite{Strassberg2019,Tolmachev2020} and electron nuclear double resonance (ENDOR),\cite{Babunts2023} and magnetic polaron formation has been reported.\cite{Beaulac2009,Rice2017}

In particular, DMS two-dimensional colloidal nanoplatelets (NPLs) are interesting because  NPLs have extraordinary optical properties compared to spherical NCs and nanorods.\cite{Diroll2023} The most studied DMS NPLs have CdSe cores and doped CdMnS shells.\cite{Delikanli2015,Murphy2016,Muckel2018,Dehnel2020,Shornikova2020acsn} In these structures the electrons and holes are confined in nonmagnetic CdSe core, and penetration of their wave functions into the CdMnS shells provides the  $sp-d$ exchange interactions. CdMnS NPLs with strong overlap of the carrier wave function with Mn$^{2+}$ ions have also been synthesized,\cite{Davis2019} and silver-doped CdSe NPLs have demonstrated magnetic behavior.\cite{Shabani2023}

Photoluminescence (PL) of excitons in DMS CdSe-based quantum dots is found only if exciton energy is below the Mn emission at 2.1 eV, which is valid only for large-size NCs.\cite{Beaulac2008nl-1} CdSe/CdMnS NPLs do not demonstrate Mn emission,\cite{Shornikova2020acsn,Delikanli2015} probably, due to a small penetration of electron and hole wavefunctions into shells, which is strong enough for $sp-d$ exchange, but not sufficient for effective energy transfer from excitons to Mn.

There is a very limited number of studies of exciton polarized emission in high magnetic fields in DMS colloidal NCs. CdSe-based quantum dots\cite{Beaulac2008nl-1,Nelson2015} and NPLs\cite{Delikanli2015,Shabani2023} have been studied in magnetic fields up to only 5 T. A rapid increase and saturation of degree of circular polarization (DCP) are reported. Among other materials, Mn doped PbS colloidal quantum dots were measured up to 30 T.\cite{Turyanska2014} We have previously reported on magneto-optical studies of DMS NPLs in magnetic fields up to 15 T.\cite{Shornikova2020acsn} A rapid increase of DCP followed by a slow decrease was observed.

In this paper, we study the magneto-optical properties of core/shell CdSe/CdMnS NPLs, which arise from excitons interacting with the magnetic Mn$^{2+}$ ions. Experiments are performed in high magnetic fields up to 30 T and at cryogenic temperatures down to 4 K. A non-monotonous dependence of the degree of circular polarization (DCP) of PL on a magnetic field is found. We demonstrate that the interplay between the dark and bright exciton states is responsible for this behavior and develop a model, which qualitatively describes the experimental results.

\section{Results and Discussion}
\begin{figure*}[t!]
	\begin{center}
		\includegraphics{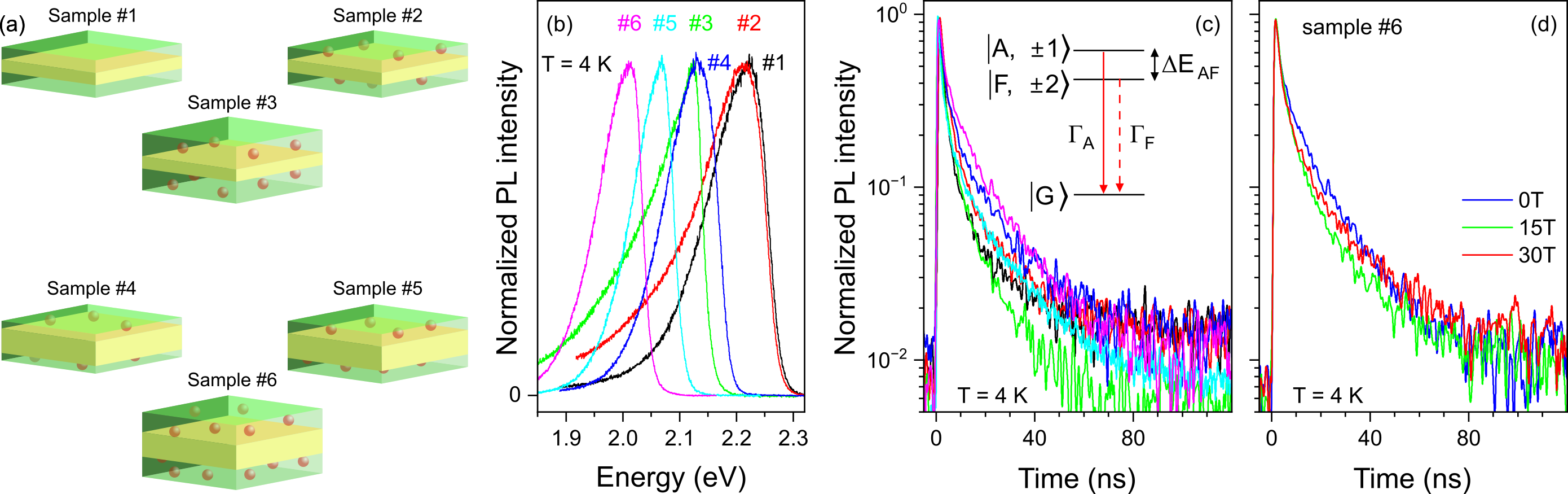}
		\caption{\label{fig1} (a) Cartoons representing  the studied NPL samples. Red balls mark Mn$^{2+}$ ions. (b) Photoluminescence spectra of the NPL samples at $T=4$ K. (c) Recombination dynamics measured at emission maxima. Color codes are same as in panel (b). Inset presents the three-level model: $\ket{A}$ and $\ket{F}$ are the bright and dark exciton states, $\ket{G}$ is the unexcited crystal state, $\Delta E_{AF}$ is the bright-dark exciton splitting, $\Gamma_A$ and $\Gamma_F$ are the radiative recombination rates of bright and dark excitons, respectively. (d) Recombination dynamics in sample \#6 measured at emission maximum in magnetic fields of 0, 15 and 30~T at $T=4$~K.}   
	\end{center}
\end{figure*}

Six CdSe/CdMnS and CdSe/CdS NPL samples were grown for this study, see Refs.\onlinecite{Delikanli2015,Delikanli2019,Shendre2019} and Methods for details. Samples \#1 -- \#3 have 2 ML (monolayer) thick CdSe cores and shells consisting of 2 ML of CdS (\#1), 2 ML of CdMnS (\#2), and 6 ML of CdMnS (\#3) cladding the core. Samples \#4~-- \#6 have 4 ML thick CdSe cores and CdMnS shells with 1 ML (\#4), 2 ML (\#5), and 3 ML (\#6) thickness. All samples have similar Mn concentration of about $x=0.004$ (0.4\%) according to our previous studies on similar samples, see Refs.\onlinecite{Shornikova2020acsn,Tolmachev2020,Babunts2023} for details. The sample parameters are given in Table~\ref{tab1}, and cartoons representing the NPL structures are shown in Fig.~\ref{fig1}a.

\begin{table*}
	\small
	\caption{Parameters of the studied CdSe/CdS and CdSe/CdMnS NPLs}
	\begin{tabular}{ |l|l|l|l|l|l|l|l| } 
		\hline
		\parbox{1.2cm}{Sample} & \parbox{1.6cm}{Core thickness} & \parbox{1.6cm}{Shell thickness} & \parbox{1.6cm}{Total thickness} & \parbox{1.8cm}{\vspace{2mm}$f_e$ ($\Delta E_c=0.3$ eV)\vspace{2mm}} & \parbox{1.8cm}{$f_h$ ($\Delta E_v=0.45$ eV)} & \parbox{1.75cm}{Mn concentration} & \parbox{2.7cm}{Emission energy ($T = 4$ K), eV}\\
		\hline
		\#1 & 2 ML & 2 ML & 6 ML  & 0.37 & 0.22 &0&2.22\\
		\#2 & 2 ML & 2 ML & 6 ML  & 0.37 & 0.22 &0.4\%&2.21\\ 
		\#3 & 2 ML & 6 ML & 14 ML & 0.58 & 0.26 &0.4\%&2.12\\
		\#4 & 4 ML & 1 ML & 6 ML  & 0.08 & 0.03 &0.4\%&2.13\\
		\#5 & 4 ML & 2 ML & 8 ML  & 0.18 & 0.06 &0.4\%&2.07\\ 
		\#6 & 4 ML & 3 ML & 10 ML & 0.24 & 0.07 & 0.4\%&2.01\\ 
		\hline
	\end{tabular}
	\label{tab1}
\end{table*}

Figure \ref{fig1}b shows PL spectra of all samples measured at $T=4$ K. Samples \#1 and \#2 have similar core and shell thickness, but sample \#2 has Mn in the shell. These two samples have very similar PL spectra with emission maxima at the energies of 2.22 and 2.21 eV, respectively. This shows that implementation of a small Mn concentration does not change the exciton emission. Emission of sample \#3 with the same core and a thicker shell is redshifted to 2.12 eV due to the penetration of electron and hole wavefuctions to the shell, which reduces the quantization energies of the electrons and holes. Samples \#4 -- \#6 have similar cores with 4 ML thickness and shells increasing from 1 to 3 ML, and their emission maxima are at 2.13, 2.07, and 2.01 eV, with smaller energies corresponding to thicker shells. All six samples have fwhm (full width at half maximum) of about 100 meV, which is provided by the inhomogeneous broadening of the NPL ensembles.

Figure \ref{fig1}c shows recombination dynamics measured at emission maxima with time resolution of 200 ps. All dynamics are quite similar, they are multiexponential with an initial fast decay on a timescale of about 3 ns followed by a slower decay. The average lifetimes estimated from three-exponential decays are from 15 to 40 ns with an error of $\pm5$ ns (Methods). 

Figure \ref{fig1}d shows recombination dynamics measured at emission maximum in sample \#6 at various magnetic fields up to 30 T. The decay remains unchanged under applied magnetic field. Additional data on emission decays in samples \#1--\#6 measured at various temperatures and magnetic fields are presented in Supporting Information \al{S1}.

The exciton recombination dynamics in our samples is weakly affected by temperature or magnetic field. We discussed this issue in similar samples in Ref.\onlinecite{Shornikova2020acsn}. In short, in CdSe-based core/shell NPLs, the exciton fine structure has five states,  with the two lowest exciton states being the dark exciton $\ket{F}$ with the angular momentum projections $\ket{\pm 2}$ and the bright exciton $\ket{A}$ with the angular momentum projections $\ket{\pm 1}$\cite{Efros1996,Shornikova2018,Biadala2014} (see inset in Fig.~\ref{fig1}c). The other three states with angular momentum projections $\ket{0^L}$, $\ket{\pm1^U}$, and $\ket{0^U}$ are separated by more than 100 meV from the $\ket{A}$ and $\ket{F}$ states and do not participate in emission. The splitting between the $\ket{A}$ and $\ket{F}$ states, $\Delta E_{AF}$, is typically about 1 meV in similar NPLs,\cite{Shornikova2020acsn,Smirnova2023} which is comparable to $k_B T$ even at cryogenic temperatures. Here $k_B$ is the Boltzmann constant. Excitons populate both the dark and bright exciton states. The dark state has higher occupation than the bright state, but its radiative recombination rate is smaller: $\Gamma_F < \Gamma_A$, and the emission is contributed by the dark and bright excitons. Here $\Gamma_F$ and $\Gamma_A$ are the radiative recombination rates of the dark and bright excitons, respectively. As a result, the PL dynamics is not sensitive to the changes of the magnetic field and temperature, as it is the case in colloidal bare CdSe NPLs and NCs with large bright-dark splittings.\cite{Shornikova2018,Shornikova2020nl,JohnstonHalperin2001,Biadala2014, Biadala2009} The magnetic field, on one hand, splits the exciton energy levels, which leads to thermal redistribution of exciton populations of these levels. On the other hand, magnetic field mixes the bright and dark exciton states accelerating the dark exciton and slowing down the bright exciton radiative rates. Our experimental accuracy of measuring radiative recombination rate is about $\pm30\%$ (Methods), and we cannot resolve small variations of the rate.

\begin{figure*}[t!]
	\begin{center}
		\includegraphics{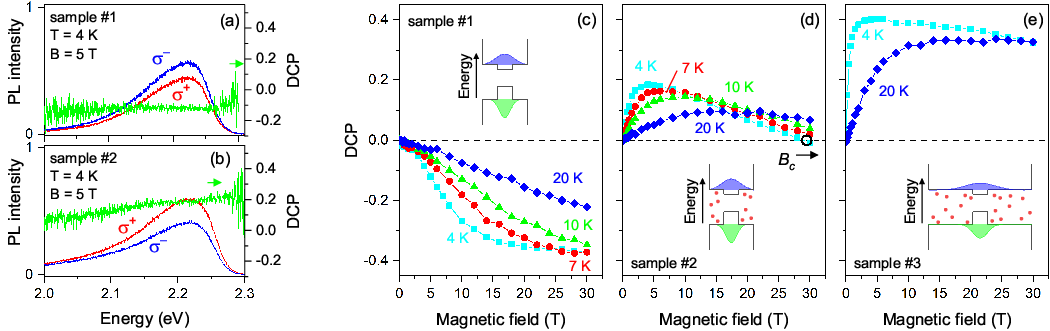}
		\caption{\label{fig2} (a,b) Spectra of the $\sigma^+$ (red) and $\sigma^-$ (blue) polarized PL in samples \#1 and \#2 measured at $T=4$ K and $B=5$ T. Left scale: Spectrally-resolved degree of circular polarization (DCP, green). (c--e) DCP measured at PL maximum as a function of magnetic field at various temperatures: (c) sample \#1, (d) sample \#2, and (e) sample \#3. Cartoons schematically show the band structures and electron and hole wave functions (blue and green contours).}   
	\end{center}
\end{figure*}

Note that the PL dynamics, which are not affected by temperature and magnetic field, could indicate trion (charged exciton) recombination. However, we are convinced that the excitons are the main emitting state in these NPLs, for two reasons: (i) exciton emission has been reported in various core/shell CdSe/CdS NPLs with thin shells,\cite{Strassberg2019,Shornikova2020acsn,Smirnova2023} (ii) trion emission would not explain the non-monotonous dependence of the DCP on a magnetic field (see below). It is possible that the trion emission also has some impact to the PL. Taking trions into account would not change the results, but complicate the modeling. Therefore, we consider only exciton emission and neglect the trions.

Figures~\ref{fig2}a and \ref{fig2}b show circularly polarized PL spectra of samples \#1 and \#2 measured at $T=4$ K in a magnetic field of $B=5$ T. The spectra are normalized so that the intensity of the sum of $\sigma^+$ and $\sigma^-$ polarized PL at emission maximum equals to unity. In the nonmagnetic sample \#1 the $\sigma^-$ polarized PL has larger intensity, while in DMS sample \#2 the $\sigma^+$ polarized PL is stronger. The DCP is calculated using the following equation:
\begin{equation}
	\textit{DCP}=\frac{I^+-I^-}{I^++I^-},
\end{equation}
where $I^+$ and $I^-$ are the intensities of $\sigma^+$ and $\sigma^-$ polarized PL, respectively. In sample \#1, the DCP is negative and equals to $-0.1$, while in sample \#2 the DCP is positive and equals to $+0.2$. In all studied samples the spectral dependence of DCP is weak or absent. 

Figures~\ref{fig2}c--e show the DCP in samples \#1~-- \#3 measured at various temperatures in magnetic fields from 0 to 30~T. In sample \#1 at 4~K the absolute DCP value increases in $B<10$~T, and starts to saturate in $B>10$~T reaching $-0.36$ in high magnetic fields. At higher temperatures, the absolute DCP value increases slower with the magnetic field, and at 20~K the DCP cannot reach the saturation. In sample \#2 at 4~K the DCP depends on the magnetic field non-monotonously. Initially, it is positive and reaches $+0.18$ at $B=5$~T. In stronger fields the DCP decreases and crosses zero at a critical magnetic field $B_c=29.5$~T. When the temperature increases, the DCP maximum value reduces and $B_c$ shifts to higher magnetic fields, which cannot be reached in our experiments. Sample \#3 has positive DCP, which reaches $+0.4$ already at $B=3$~T at 4~K, and slightly decreases in higher magnetic fields. At 20~K the DCP saturates at $+0.33$ for $B > 10$~T. 

\begin{figure*}[t!]
	\begin{center}
		\includegraphics{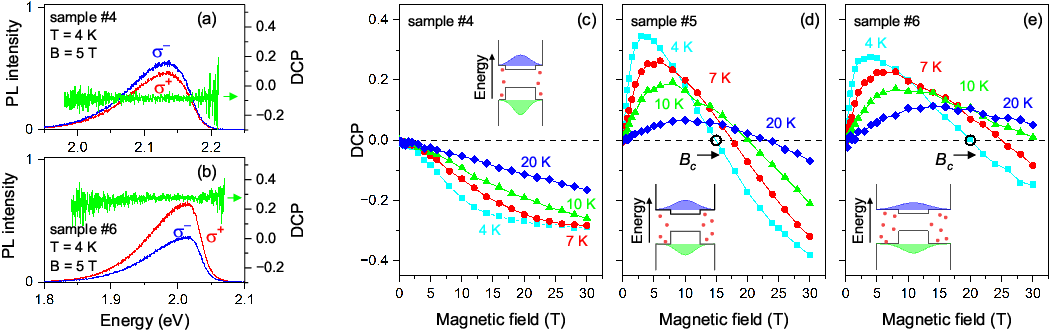}
		\caption{\label{fig3} (a,b) Spectra of the $\sigma^+$ (red) and $\sigma^-$ (blue) polarized PL in samples \#4 and \#6 measured at $T=4$ K and $B=5$ T. Left scale: Spectrally-resolved degree of circular polarization (DCP, green). (c--e) DCP measured at PL maximum as a function of magnetic field at various temperatures: (c) sample \#4, (d) sample \#5, and (e) sample \#6. Cartoons schematically show the band structures and electron and hole wave functions (blue and green contours).}   
	\end{center}
\end{figure*}

Qualitatively similar behavior is found in samples \#4 -- \#6. Figures~\ref{fig3}a,b show polarized PL spectra of samples \#4 and \#6 at $B=5$~T. In sample \#4, the DCP is negative and equals to $-0.08$, while in sample \#6 it is positive and equals to $+0.28$. Figures~\ref{fig3}c--e show DCP in samples \#4 -- \#6 at various temperatures measured in magnetic fields from 0 to 30~T. In sample \#4 at 4 K, the DCP monotonously reaches the saturation value of $-0.29$ at $B=30$~T. At higher temperatures, the absolute DCP value increases slower with a field growth and at 20~K the DCP cannot reach the saturation. Surprisingly, despite the presence of Mn$^{2+}$ in sample \#4, its DCP is negative and shows behavior similar to the nonmagnetic sample \#1, compare with Fig.~\ref{fig2}c. It is due to the very weak interaction of excitons with the Mn$^{2+}$ ions in this sample with thin CdMnS core. In samples \#5 and \#6, the DCP is non-monotonous. In sample \#5 at 4~K it reaches $+0.35$ at $B=3.5$~T, crosses zero at $B_c=15$~T, and decreased to $-0.38$ at $B=30$~T, still not saturating. At higher temperatures $B_c$ shifts to higher fields reaching $B_c=24$~T at 20~K. In sample \#6 the behavior is similar, but $B_c$ is slightly higher: $B_c=20$~T at 4~K, and more than 30~T at 20~K.

The DCP dependence on magnetic field in samples \#1 and \#4 is typical for CdSe-based nonmagnetic colloidal NCs,\cite{JohnstonHalperin2001,Qiang2021,GranadosdelAguila2017} but the DCP observed in other samples, is unusual. There are two experimental facts, which have to be explained. First, a non-monotonous DCP dependence on a magnetic field, and second, an increase of $B_c$ with increasing temperature.

Three mechanisms could be responsible for that: (i) interplay between the intrinsic and exchange Zeeman splittings of excitons or trions, (ii) spin-dependent recombination of dark excitons, and (iii) interplay between bright and dark exciton states emitting in $\sigma^+$ and $\sigma^-$ polarizations, respectively. Let us consider these mechanisms. We use the conventional approach for diluted magnetic semiconductors, which is based on consideration of the exchange interaction of an exciton with the Mn$^{2+}$ ions within the mean field approximation.\cite{Furdyna1988book,Kossut2010} Recently, we have successfully implemented this approach to describe the DCP in CdSe/CdMnS NPLs.~\cite{Shornikova2020acsn,Delikanli2015}.

\textbf{(i) Interplay between the intrinsic and exchange Zeeman splittings of excitons or trions}. The exchange interaction of electron band states with localized magnetic ions results in  the giant Zeeman splitting. The Zeeman splitting of electrons, holes or excitons in DMS materials has two contributions:
\begin{equation}
	\Delta E_Z(B)=\Delta E_{Z}^{\rm intr}(B)+\Delta E_{Z}^{\rm exch}(B).
\end{equation}
Here $\Delta E_Z^{\rm intr}(B)$ is the intrinsic Zeeman splitting characteristic for nonmagnetic NCs. In the presence of an external magnetic field $\vec{\boldsymbol{B}}$ a particle with a magnetic moment $\vec{\boldsymbol{\mu}}$ acquires an energy $\Delta E_Z^{\rm intr}(B)=-\vec{\boldsymbol{\mu}}\cdot\vec{\boldsymbol{B}}$. The magnetic moment can be expressed as  $\vec{\boldsymbol{\mu}}=g\mu_B\vec{\boldsymbol{J}}$, where $g$ is the $g$-factor, $\mu_B$ is the Bohr magneton, and $\vec{\boldsymbol{J}}$ is the angular momentum of the particle. The energies acquired by, for example, bright exciton states with the angular momentum projections $J_Z=\ket{-1}$ and $J_Z=\ket{+1}$ are different, and a Zeeman splitting between these two states appears. $\Delta E_{Z}^{\rm exch}(B)$ represents the exchange energy due to the interaction of excitons or charge carriers with magnetic Mn$^{2+}$ ions.

The Zeeman splitting of electrons in CdSe/CdMnS NPLs is:
\begin{equation}
	\begin{gathered}
		\Delta E_{Z,e}(B)=\Delta E_{Z,e}^{\rm intr}(B)+\Delta E_{Z,e}^{\rm exch}(B)= \\
		 =g_e \mu_B B-\langle S_{\rm Mn}(B) \rangle x N_0 \alpha f_e,
	\end{gathered}
	\label{eq:3}
\end{equation}
and of holes:
\begin{equation}
	\begin{gathered}
	\Delta E_{Z,h}(B)=\Delta E_{Z,h}^{\rm intr}(B)+\Delta E_{Z,h}^{\rm exch}(B)= \\
	=-3g_h \mu_B B-\langle S_{\rm Mn}(B) \rangle x N_0 \beta f_h.
	\label{eq:4}
	\end{gathered}
\end{equation}
Here $g_e$ and $g_h$ are the electron and hole $g$-factors, respectively, $\langle S_{\rm Mn}(B) \rangle$ is the mean spin of Mn$^{2+}$ ions, $x$ is the Mn concentration, $N_0$ is the number of cations per unit volume, $\alpha$ and $\beta$ are the \textit{s}--\textit{d} and \textit{p}--\textit{d} exchange constants, respectively. $f_e$ and $f_h$ are the probabilities to find an electron and a hole in the shells, respectively. Here we assume that the heavy hole is the ground hole state, which is true for CdSe-based NPLs.\cite{Shornikova2018,Biadala2014} We use the definition of the hole $g$-factor sign that is commonly used for colloidal NCs.\cite{Shornikova2018nl,Efros2003}

In the studied samples the Mn$^{2+}$ ions are located in the shells, and the penetration of the electron and hole wave functions into the shells controls the exchange interaction. The penetration can be calculated using a particle in a box model, see Supporting Information~\al{S3}. Since the conduction and valence band offsets between CdSe and CdS are not precisely known, the probabilities to find an electron and a hole in the shell can be estimated with certain tolerance. The values of  $f_e$ and $f_h$, which we used for calculations of the DCP, are given in Table~\ref{tab1}. The electron and hole wave functions are shown schematically in insets of Figs.~\ref{fig2}c--e and \ref{fig3}c--e.

According to eqs. \eqref{eq:3} and \eqref{eq:4}, the exchange part of the carrier Zeeman splitting depends on the mean spin of the Mn$^{2+}$ ions $\langle S_{\rm Mn} (B)\rangle$. The dependence of $\langle S_{\rm Mn} (B)\rangle$ on magnetic field is described by the modified Brillouin function for spin $5/2$, see Supporting Information \al{S4}. An example of calculated intrinsic and exchange electron and hole Zeeman splittings at various temperatures is shown in Figs.~\ref{fig:Sintr-exch_e} and \ref{fig:Sintr-exch_h}. The exchange part increases rapidly in low magnetic fields, and saturates for $B>10$~T, while the intrinsic part increases linearly with the magnetic field. This means that, if the intrinsic and exchange splittings have opposite signs, there is a finite magnetic field $B_c$, in which $\Delta E_Z^{\rm intr}(B_c)+\Delta E_Z^{\rm exch}(B_c)=0$. In this field the Zeeman splitting reverses the sign, and so does the DCP.

The competition between the intrinsic and exchange splittings can explain the non-monotonous dependence of the DCP on magnetic field. This is valid for negatively charged trions and excitons, but not for positively charged trions, which are not expected to demonstrate a non-monotonous DCP, since their intrinsic and exchange splittings have the same signs. However, this competition does not explain the shift of the critical magnetic field $B_c$ to higher fields with increasing temperature. Indeed, the average spin of Mn${}^{2+}$ ions decreases with the temperature increase and, consequently, the exchange splitting reduces. This means that the intrinsic splitting compensates the exchange splitting at lower field, if the temperature increases, or, in other words, $B_c$ decreases with the temperature increase.  This is valid for any two-level system: negatively charged trions, dark excitons, and bright excitons, see Supporting Information \al{S4} and \al{S5} for more details. 

\textbf{(ii) Spin-dependent dark exciton  recombination.} A non-monotonous dependence of the DCP on magnetic field, with the sign reversal from positive to negative, has been observed in nonmagnetic bare core CdSe NPLs.\cite{Shornikova2020nn} The effect is caused by the spin-dependent recombination of the dark exciton assisted by flips of the surface spins. The surface spins are polarized by magnetic field, making the emission of excitons with angular momentum projection $\ket{+2}$, which is $\sigma^+$ polarized, more favorable than the emission of excitons with angular momentum projection $\ket{-2}$, which is $\sigma^-$ polarized. In low magnetic fields, this effect can overcome the polarization caused by thermal population of the Zeeman levels. In other words, while the exciton state with momentum projection $\ket{-2}$ has lower energy and higher occupation, the emission from the upper $\ket{+2}$ state is stronger due to the spin-dependent recombination \latin{via} surface spins. When magnetic field increases, the Zeeman splitting grows, and the occupation of the $\ket{-2}$ state increases. This is favorable for the $\sigma^-$ polarized emission. Therefore, the DCP is positive in low magnetic fields, and reverses the sign to negative in high magnetic fields. For this mechanism under certain conditions, $B_c$ increases with increasing temperature.

In core/shell NPLs, which are used in the present study, electron and hole wave functions are mostly localized in the cores and partly penetrate into the shells, but their amplitudes at the surface are very small compared to the bare core NPLs. Moreover, the surface spin-dependent recombination is not connected to the presence of Mn, and would manifest itself in the nonmagnetic sample \#1, but a positive DCP is observed only in DMS samples. Therefore, we exclude the spin-dependent recombination assisted by surface spins from possible explanations of the non-monotonous DCP.  

The localized spins of magnetic Mn$^{2+}$ ions can act for spin-dependent recombination similar to the surface spins. We modeled a situation, when only the dark exciton is involved in recombination, and the increase of $B_c$ with increasing temperature is explained by strong Mn-assisted spin-dependent recombination process, see Supporting Information \al{S7} for more details. However, this hypothesis requires extremely strong Mn-assisted recombination with a rate four times faster than $\Gamma_F$. This would result in much faster PL dynamics in DMS NPLs compared to nonmagnetic sample \#1, which is not observed experimentally (Fig.~\ref{fig1}c). We also did not observe such strong recombination acceleration in our previous studies on similar NPLs.\cite{Shornikova2020acsn}

The requirement for very strong Mn-assisted recombination is partially due to the fact that recombination of dark excitons not assisted by Mn would lead to $B_c$ shifting to lower fields with increasing temperature (see (i)). The spin-dependent recombination should be very strong to compensate this effect.

With the arguments given above, we exclude the spin-dependent recombination as an explanation of the increase of $B_c$ with increasing temperature.

\textbf{(iii) Interplay between bright and dark exciton states emitting in $\sigma^+$ and $\sigma^-$ polarizations, respectively.} The bright exciton Zeeman splitting is calculated using the electron and hole splittings:
\begin{equation}
	\Delta E_{Z,XA}(B)= g_{XA}\mu_B B+ \Delta E_{Z,XA}^{\rm exch}(B),    %\equiv g_{XA}^{\rm eff}\mu_B B,
	\label{eq:E_XA}
\end{equation}
where $g_{XA}=-g_e-3g_h$ is the bright exciton $g$-factor,\cite{Efros1996,Semina2021} and $\Delta E_{Z,XA}^{\rm exch}(B)=-\Delta E_{Z,e}^{\rm exch}+\Delta E_{Z,h}^{\rm exch}$ is the exchange term. Similarly, the dark exciton Zeeman splitting reads as:
\begin{equation}
	\Delta E_{Z, XF}(B)= g_{XF}\mu_B B+ \Delta E_{Z,XF}^{\rm exch}(B),  %\equiv g_{XA}^{\rm eff}\mu_B B,
	\label{eq:E_XF}
\end{equation}
where $g_{XF}=g_e-3g_h$ is the dark exciton $g$-factor,\cite{Efros1996,Semina2021} and $\Delta E_{Z,XF}^{\rm exch}(B)=\Delta E_{Z,e}^{\rm exch}+\Delta E_{Z,h}^{\rm exch}$ is the exchange term.

%Note, that the exciton exchange interaction is controlled by the exchange with Mn$^{2+}$ spins of both electrons and holes composing the excitons. However, the exchange interaction of electrons is smaller than that of holes and has the same sign as the intrinsic electron splitting. Therefore, the holes play the dominant role in exchange interaction of excitons with the magnetic ions in CdSe/CdMnS NPLs.\cite{Shornikova2020acsn}

\begin{figure*}[t!]
	\begin{center}
		\includegraphics{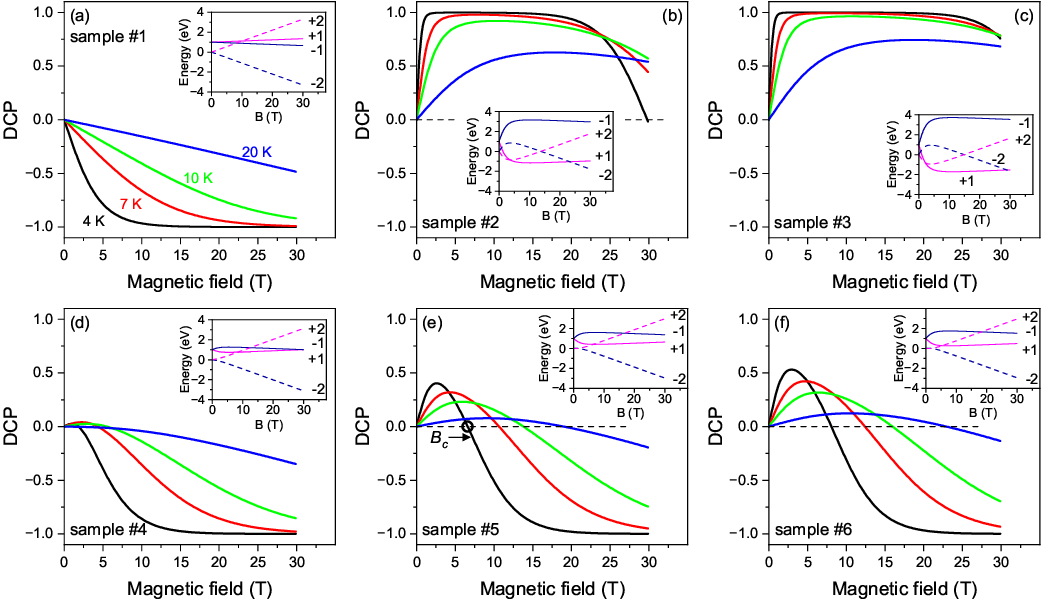}
		\caption{\label{fig4} Calculated degree of circular polarization (DCP) of excitons as a function of magnetic field at various temperatures for samples \#1--\#6 with parameters given in text. Insets show calculated exciton energies at $T=4$~K, the $\sigma^-$ emitting states are dark blue, and the $\sigma^+$ emitting states are pink.}   
	\end{center}
\end{figure*}

Figure~\ref{fig4} shows the modeled DCP at various temperatures in all samples. The insets show corresponding energy levels of the bright and dark excitons as functions of the magnetic field at $T=4$ K. The following parameters are used: $g_e=1.7$,\cite{Shornikova2020acsn,Kudlacik2020,Feng2020} $g_h=-0.7$,\cite{Shornikova2020nl} $N_0\alpha=0.22$~eV and $N_0\beta=-1.8$~eV \cite{Kossut2010}, $\Delta E_{AF}=1$ meV,\cite{Shornikova2020acsn,Smirnova2023} $f_e$ and $f_h$ listed in Table~\ref{tab1}, $\Gamma_A=0.1$  ns$^{-1}$, $\Gamma_F=0.01$ ns$^{-1}$ (see Supporting Information \al{S1}). Details of the modeling can be found in Supporting Information \al{S4, S5}. 

The intrinsic dark and bright exciton $g$-factors are positive ($+3.8$ and $+0.4$, respectively). Therefore, in the nonmagnetic sample \#1 the state with $\ket{-2}$ angular momentum projection has lower energy than the $\ket{+2}$ state, and the $\ket{-1}$ state has lower energy than the $\ket{+1}$ state (inset Fig.~\ref{fig4}a). The intrinsic dark exciton splitting is larger than the intrinsic bright exciton splitting, compare the energy separation between the $\ket{-2}$ and $\ket{+2}$ states with the splitting between the $\ket{-1}$ and $\ket{+1}$ states. We neglect the anticrossing of exciton energy levels in our calculations, because we observe no indication of it (sharp resonances in dependences of total PL intensity and DCP on a magnetic field\cite{Ivchenko2018}), see Supporting Information \al{S6} for more details.

The PL from sample~\#1 is $\sigma^-$ polarized in a magnetic field (Fig.~\ref{fig4}a). At $T=4$~K it is mostly contributed  by emission from the $\ket{-2}$ exciton state.  At higher temperatures the polarization degree decreases, because the other states are thermally populated and participate in emission. Here in modeling we assume the Boltzmann exciton distribution. Note, that we do not take into account the depolarizing factors, so that the absolute value of DCP at low temperatures and in high magnetic fields reaches 1 (\latin{i.e.} 100\%). In real NCs, various factors can reduce the polarization: orientation of the NPL quantization axis arbitrary to the magnetic field,\cite{Shornikova2020nn,Shornikova2018nl} dynamical factor,\cite{Liu2014} phonon-assisted recombination,\cite{Rodina2016,Qiang2021} and linearly polarized emission due to the NPL anisotropic shape.\cite{Feng2018}  The absolute values of the DCP experimentally measured in this study do not exceed 0.4. Therefore, our calculations reproduce the shape, but not the absolute values.

Sample \#2 (Fig.~\ref{fig4}b) has the same design as sample \#1, but the shell is doped with Mn. The exchange interaction alters the energies of the exciton states in a magnetic field. The bright exciton $\ket{+1}$ state is lower than the $\ket{-1}$ state, and the splitting between them quickly increases with growing field and saturates at about 5~T. For the dark exciton the exchange Zeeman splitting also inverts the $\ket{\pm2}$ states in low magnetic fields, so that the $\ket{+2}$ state has lower energy. However, in magnetic fields above 14 T, the intrinsic splitting of dark exciton dominates, and the $\ket{-2}$ dark exciton state is below the $\ket{+2}$ state. Above 23 T the lowest exciton state is $\ket{-2}$. The corresponding DCP (black line in Fig.~\ref{fig4}b) is positive and reverses sign at 30~T. The magnetic field, in which the sign is reversed, is determined by the competition between the emission of the thermally populated $\ket{-2}$ and $\ket{+1}$ states. At higher temperatures, the  $\ket{+1}$ level is populated in magnetic fields up to 30 T, and the DCP is positive (red, green and blue lines in Fig.~\ref{fig4}b).

Sample \#3 (Fig.~\ref{fig4}c) has thicker shell and larger exchange energy. The $\ket{+2}$ state below 2~T and $\ket{+1}$ state above 2 T are the lowest in energy, and the emission is $\sigma^+$ polarized. At 4~K the DCP saturates already at 1 T and stays at this nearly saturated level up to 30 T. At 20~K the DCP  increases in magnetic fields 0--15 T and stays about constand above 15 T.

Sample~\#4 (Fig.~\ref{fig4}d) has 4 ML thick core and 1 ML thick CdMnS shells. Due to very small penetration of the electron and hole wavefunctions into the shell (Table~\ref{tab1}), the exchange Zeeman splittings are very small. In bright exciton the exchange splitting slightly overcomes the intrinsic splitting, and the $\ket{+1}$ state has slightly lower energy than the $\ket{-1}$ state. In dark exciton the intrinsic splitting overcomes the exchange splitting, and the lowest dark exciton state is $\ket{-2}$. As a result, the DCP is positive in low magnetic fields due to the dominating emiission from the $\ket{+1}$ state, and reverses the sign already in 2 T at 4 K. In experiment, we observed negative DCP starting from 0~T (compare with Fig.~\ref{fig3}c).

Samples~\#5 and \#6 (Figs.~\ref{fig4}e,f) have 2 and 3 ML thick shells, respectively. The penetrations of the electron and hole wave functions into the shells are stronger than in sample \#4, which increases the exchange with Mn$^{2+}$. As we discussed above, the bright exciton has about an order of magnitude smaller intrinsic Zeeman splitting than the dark exciton, because of the different $g$-factors. Additionally, the bright exciton has larger exchange Zeeman splitting than the dark exciton, because the electron exchange energy $\Delta E_{Z,e}^{\rm exch}$ enters eqs~\eqref{eq:E_XA} and \eqref{eq:E_XF} with opposite signs. As a result, the exchange with Mn$^{2+}$ spins affects the bright exciton stronger than the dark exciton. The $\ket{+1}$ bright exciton state is always below the $\ket{-1}$ state, while in dark exciton the intrinsic splitting overcomes the exchange at 1.5 and 2 T in samples \#5 and \#6, respectively, at 4 K. In low magnetic fields at 4 K, the emission is $\sigma^+$ polarized, because the $\ket{+1}$ bright exciton level is populated, and it has faster recombination rate. With increasing magnetic field, the splitting between the lowest $\ket{-2}$ and the upper $\ket{+1}$ level increases, and above about 6 T in sample \#5 (8 T in sample \#6), when only the $\ket{-2}$ level is populated, the DCP is $\sigma^-$ polarized. At higher temperatures, the stronger magnetic field is needed to depopulate the $\ket{+1}$ level, and the critical magnetic field $B_c$ shifts to higher fields.

Comparison of the modeling with experimental data indicates that the interplay between the bright and dark exciton states emitting in opposite polarizations can explain both a nonmonotounous dependence of DCP on a magnetic field and an increase of the critical magnetic field $B_c$ with increasing temperature.

We believe that a similar effect can be observed in various DMS and some nonmagnetic colloidal NCs. We found a similar behavior in CdSe/CdZnMnS NPLs, see Supporting Information \al{S2}. According to our calculations, NCs with a small bright-dark splitting and opposite signs of $\Delta E_{Z,XA}$ and $\Delta E_{Z,XF}$ are good candidates to observe this bright-dark exciton interplay (Supporting Information \al{S8}). Bright-dark splitting $\Delta E_{AF}$ is an important parameter, which should not exceed several meV, otherwise the temperature population of the bright state is too low. We predict that a similar interplay can be observed in nonmagnetic NCs with $g_e=1.7$ and $g_h=-0.2$, and large-size DMS quantum dots, where both the exciton emission is not quenched and $\Delta E_{AF}$ is small.

\section{Conclusions}

In summary, magneto-optical experiments on CdSe/CdMnS nanoplatelets reveal strong influence of the exchange of excitons with Mn$^{2+}$ ions on the circular polarization of PL. The competition between the bright and dark exciton emission leads to a non-monotonous dependence of the degree of circular polarization on the magnetic field strength. %The polarization is positive in low magnetic fields, reverses sign at about 15 T and is negative in higher fields. The critical magnetic field, in which the polarization reverses the sign, increases with increasing temperature. %We explain it by competition between the bright and dark exciton emission.
The electron and hole exchange interactions with Mn$^{2+}$ ions can be tuned by modifying the composition of nanoplatelets (material and thickness of core and shell layers, number of shells, Mn concentration). This can be done with high precision since nanoplatelets are grown layer by layer. This makes them unique among the whole family of colloidal DMS nanocrystals. This flexibility can be used to reveal new spin-dependent effects in DMS colloidal nanocrystals.

\section{Methods}

\textbf{Sample preparation.} \textit{Chemicals:} Cadmium acetate dihydrate ($\text{Cd(OAc)}_2\cdot 2\text{H}_2\text{O}$), trioctylamine (TOA), oleylamine (OLA), N-methylformamide (NMF), ammonium sulfide solution ($40-48$ wt. \% in water), trioctylphosphine (TOP), oleic acid (OA), hexane, acetonitrile, toluene and manganese(II) acetate were bought from Sigma-aldrich.

\textit{Synthesis of 2 ML CdSe Nanoplatelets:} The synthesis of the NPLs was performed according to a previously reported method.~\cite{Delikanli2019} The mixture of 860 mg $\text{Cd(OAc)}_2\cdot 2\text{H}_2\text{O}$, 1 mL of OA, and 60 mL of TOA was degassed for 1 h at room temperature. Then, it was heated to 115$^\circ$C under argon flow. When the temperature reached 115$^\circ$C, 1 mL of 1M TOP-Se was injected swiftly and the mixture was kept at 115$^\circ$C for 2 h. After that, the solution was cooled down to room temperature and centrifuged after addition of ethanol and hexane. The precipitated NPLs were dispersed in hexane.

\textit{Synthesis of CdSe/CdMnS and CdSe/CdZnMnS core/shell NPLs:} Here we used a modified procedure of the c-ALD recipe reported previously.~\cite{Shendre2019} 2 ML CdSe NPLs were dispersed in 1 mL hexane and 5 mL of NMF with 40 $\mu$L of $40-48\%$ aqueous solution of ammonium sulfide -- as sulfur shell growth precursor -- was added on top of the NPL dispersion and stirred for 2 min. Then, the reaction was stopped by addition of acetonitrile and excess toluene and the mixture was precipitated \textit{via} centrifugation. The precipitate was redispersed in NMF and precipitated again after addition of acetonitrile and toluene to remove the unreacted precursor. Finally, the NPLs were dispersed in 4 mL of NMF. The cation precursor solution consists of $\text{Mn(OAc)}_2$ and $\text{Cd(OAc)}_2\cdot 2\text{H}_2\text{O}$ in NMF for the deposition of CdMnS shell to obtain CdSe/CdMnS core/shell NPLs, and $\text{Mn(OAc)}_2$, $\text{Cd(OAc)}_2\cdot 2\text{H}_2\text{O}$ and $\text{Zn(OAc)}_2\cdot 2\text{H}_2\text{O}$ in NMF for the deposition of CdZnMnS shell to obtain  CdSe/CdZnMnS core/shell NPLs. For the cation deposition step, 1 mL of cation precursor mixture was added to the NPL dispersion and it was stirred for 45 min in a nitrogen filled glovebox. Then, the reaction was stopped by addition of excess toluene and the mixture was precipitated \textit{via} centrifugation and dispersed in NMF. The same cleaning step was repeated twice more to remove the excess precursors. To increase the number of shells, the steps explained above were repeated until the desired shell thickness was achieved. Lastly, 5 mL of hexane and 100 $\mu$L of OLA were added on top of the precipitated NPLs after achievement of the desired shell thickness and the mixture was stirred overnight. To remove the excess ligands, the dispersion of NPLs was precipitated by addition of ethanol, redispersed and kept in hexane for further usage. The doping levels were obtained using ICP-MS measurements and by taking into account the 2D planar geometry of the NPLs.

Samples \#1--\#3 had cores synthesized in the same batch. Samples \#2 and \#3 were synthesized in the same c-ALD synthesis. This means that, after growing 2 ML shells on each side of the cores, a portion of the solution was taken to finish the synthesis and produce sample \#2, while the rest was used to grow 4 ML thicker shells and produce sample \#3.

Similarly, samples \#4--\#6 had the same cores and were synthesized together. After growing 1 ML shell on each side of CdSe cores, a portion of NPLs was taken to produce sample \#4. The rest was used to grow an additional shell of 1 ML, and after this procedure, a portion of NPLs was taken to produce sample \#5. The rest was taken to repeat a growth of 1 ML shell and used to produce sample \#6.

\textbf{Experiments in magnetic fields up to 30~T.} The measurements were performed in High Field Magnet Laboratory (HFML), Nijmegen. Samples were mounted on a titanium sample holder on a top of a three-axis piezo-positioner. The sample stage was placed in an optical probe, made of carbon and titanium to minimize possible displacements at high magnetic fields. Laser light was focused on the sample by a singlet objective (2~mm working distance). The same lens was used to collect the PL emission and direct it to the detection setup (backscattering geometry). The optical probe was mounted inside a liquid helium bath cryostat (4.2~K) inserted in a 50~mm bore Florida-Bitter electromagnet with a maximum $dc$ magnetic field strength of 31~T. Experiments were performed in Faraday geometry (light excitation and detection parallel to the magnetic field direction).

The PL was excited by a diode-laser operating at 405~nm (photon energy 3.06~eV) in a cw or pulsed mode (pulse duration 40 ps). The PL was dispersed with a 0.3 m spectrometer and detected either by a liquid-nitrogen-cooled charge-coupled-device (CCD) camera or by a Si avalanche photodiode (APD) connected to a conventional time-correlated single-photon counting system. The instrumental response time was about 200~ps. The PL circular polarization degree was analyzed by a combination of a quarter-wave plate and a linear polarizer. The duration of the time-resolved measurements was limited by 1 minute, which the magnet could stay in high magnetic fields. For this reason, the measured decay rates have low accuracy with an error of $\pm5$ ns.

\textbf{Lifetime measurements.} The recombination dynamics at temperatures down to 2~K were measured in TU Dortmund, see Fig. S1. The PL was excited by a diode-laser operating at 405~nm (photon energy 3.06~eV) in a pulsed mode (pulse duration 40 ps). The PL was dispersed with a 0.5 m spectrometer and detected by a Si avalanche photodiode (APD) connected to a conventional time-correlated single-photon counting system. The instrumental response time was about 200~ps.

\vspace{0.5cm}
\textbf{AUTHOR INFORMATION}\\
Corresponding Authors:\\
Email: elena.shornikova@tu-dortmund.de\\ 
Email: dmitri.yakovlev@tu-dortmund.de\\

\textbf{ORCID}\\
Elena V. Shornikova: 0000-0002-6866-9013 \\
Dmitri R. Yakovlev: 0000-0001-7349-2745 \\
Danil~O.~Tolmachev:  0000-0002-7098-8515 \\
Mikhail A. Prosnikov: 0000-0002-7107-570X\\
Peter C. M. Christianen:  0000-0002-6361-1605 \\
Sushant Shendre: 0000-0001-8586-7145\\
Furkan Isik: 0000-0001-5881-5438\\
Savas Delikanli: 0000-0002-0613-8014 \\
Hilmi  Volkan Demir:  0000-0003-1793-112X  \\
Manfred~Bayer:  0000-0002-0893-5949\\

\textbf{ASSOCIATED CONTENT}

The authors declare no competing financial interests.

\begin{acknowledgement}
Authors thank A. V. Rodina and A. A. Golovatenko for discussions. The work of E.V.S. was funded by the Deutsche Forschungsgemeinschaft (DFG, German Research Foundation) – Project No.: 462009643. D.R.Y. and M.B. acknowledge support of the Deutsche Forschungsgemeinschaft in the frame of the International Collaborative Research Center TRR 160 (Project B1). The support of HFML-RU/NWO-I, a member of the European Magnetic Field Laboratory (EMFL), is acknowledged. H.V.D. gratefully acknowledges financial support in part from the Singapore Agency for Science, Technology and Research (A*STAR) MTC program under grant number M21J9b0085, Ministry of Education, Singapore, under its Academic Research Fund Tier 1 (MOE-RG62/20), and in part from TUBITAK 121C266 and 20AG001. H.V.D. also acknowledges support from TUBA and TUBITAK 2247-A National Leader Researchers Program (121C266).

\end{acknowledgement}

\begin{suppinfo}
Additional experimental data: Time-resolved measurements, data on additional CdSe/CdZnMnS samples. Modeling details, comments. 
\end{suppinfo}

\clearpage

%%%%%%%%%% Merge with supplemental materials %%%%%%%%%%
%%%%%%%%%% Prefix a "S" to all equations, figures, tables and reset the counter %%%%%%%%%%
\setcounter{equation}{0}
\setcounter{figure}{0}
\setcounter{table}{0}
\setcounter{page}{1}
%\makeatletter
\renewcommand{\theequation}{S\arabic{equation}}
\renewcommand{\thefigure}{S\arabic{figure}}
\renewcommand{\bibnumfmt}[1]{[S#1]}
\renewcommand{\citenumfont}[1]{S#1}
\renewcommand{\thetable}{S\arabic{table}}
\renewcommand{\thepage}{S\arabic{page}}
%%%%%%%%%% Prefix a "S" to all equations, figures, tables and reset the counter %%%%%%%%%%

%\section{Supporting Information}
\onecolumn

\begin{center}
	\textbf{\large Supporting Information:}
	
	\vspace{3mm}	
	\textbf{\large Bright-Dark Exciton Interplay Evidenced by Spin Polarization in CdSe/CdMnS Nanoplatelets}
	
	\vspace{3mm}
	
	{Elena~V.~Shornikova,$^1$ Dmitri~R.~Yakovlev,$^1$ Danil~O.~Tolmachev,$^1$ Mikhail A. Prosnikov,$^2$ Peter C. M. Christianen,$^2$ Sushant Shendre,$^3$ Furkan Isik,$^3$ Savas Delikanli,$^{3,4}$ Hilmi Volkan Demir,$^{3,4}$ Manfred~Bayer$^{1}$}
\end{center}
\vspace{3mm}

{\small \noindent$^1$Experimentelle Physik 2, Technische Universit{\"a}t Dortmund, 44227 Dortmund, Germany
	
\noindent$^2$High Field Magnet Laboratory (HFML-EMFL), Radboud University, 6525 ED Nijmegen,
The Netherlands
	
\noindent$^3$}LUMINOUS! Center of Excellence for Semiconductor Lighting and Displays, School of Electrical and Electronic Engineering, School of Physical and Materials Sciences, Nanyang Technological University, 639798 Singapore

\noindent$^4$Department of Electrical and Electronics Engineering, Department of Physics, UNAM -- Institute of Materials Science and Nanotechnology, Bilkent University, 06800 Ankara, Turkey

\vspace{15mm}

\textbf{\large S1. Additional data for samples \#1 -- \#6}

\begin{figure*}[h!]
	\begin{center}
		\includegraphics{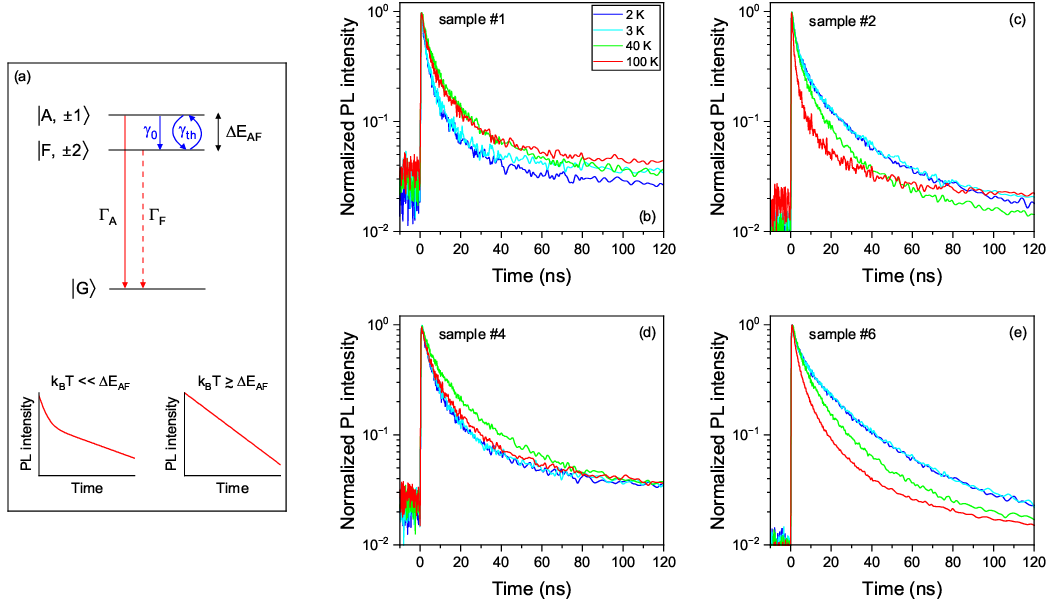}
		\caption{\label{fig:s1}\small (a) Top: three-level model. $\ket{A}$ and $\ket{F}$ are the bright and dark exciton states, $\ket{G}$ is the unexcited crystal state, $\gamma_0$ is the zero temperature relaxation rate, $\gamma_{th}$ is the thermal-activated rate for the reverse process, $\Gamma_A$ and $\Gamma_F$ are the radiative recombination rates of bright and dark excitons, respectively. Bottom left: at low temperatures when $k_B T \ll \Delta E_{AF}$ the PL dynamics is biexponential. The short decay time corresponds to the spin relaxation from the bright to the dark state, and the long decay time corresponds to the dark exciton recombination. Bottom right: at high temperatures when $k_B T \gtrsim \Delta E_{AF}$ the PL dynamics is monoexponential. (b)--(e) PL dynamics at various temperatures from 2 to 100 K in samples (b) \#1, (c) \#2, (d) \#4, and (e) \#6.}
	\end{center}
\end{figure*}

In CdSe-based NPLs the lowest exciton state, $\ket{F}$, has angular momentum projections $\ket{\pm 2}$ (Figure~\ref{fig:s1}a). The emission of this state is forbidden in electric-dipole approximation, for this reason it is called a dark exciton. The optical transitions from the upper-lying $\ket{A}$ state with angular momentum projections $\ket{\pm 1}$ are allowed, and this state is called a bright exciton. The energy splitting between these two states, $\Delta E_{AF}$, is the bright-dark splitting. The radiative recombination rates of the dark and bright exciton states are $\Gamma_F$ and $\Gamma_A$, respectively. The spin relaxation rates are: the zero-temperature relaxation rate  $\gamma_0$, and the thermal-activated rate for the reversed process $\gamma_{th}=\gamma_0N_{B}$, where $N_{B} = 1/ \left[ \exp {(\Delta E_{AF} / k_BT)} -1 \right]$ is the Bose-Einstein phonon occupation, $k_B$ is the Boltzmann constant.

In the frame of this three-level model, the exciton dynamics are commonly described as the following. At low temperatures when $k_B T \ll \Delta E_{AF}$, the decays are bi-exponential with the short decay rate $\Gamma_S$ corresponding to the spin relaxation from the bright to the dark state, and the long decay rate $\Gamma_L$ corresponding to the dark exciton recombination. With increasing temperature, the upper bright state is thermally populated, which accelerates the decay, and at $k_B T \gtrsim \Delta E_{AF}$, the decay is monoexponential, see cartoons on the bottom Fig.~\ref{fig:s1}a.

The dependence of the decay rates on temperature can be described as:\cite{Shornikova2018_s}
\begin{equation}
		\Gamma_{S,L}(T) = \frac{1}{2} \left[ \Gamma_A +\Gamma_F+\gamma_0 \coth\left( \frac{\Delta E_{AF}}{2k_B T} \right) \pm \sqrt{{\left( \Gamma_A -\Gamma_F+\gamma_0 \right)}^2+\gamma_0^2 \sinh^{-2}\left(\frac{\Delta E_{AF}}{2k_B T} \right)} \right] .
	\label{eq:s1}
\end{equation}
Here the sign ``$+$'' before the square root corresponds to $\Gamma_S$ (short decay rate) and the sign ``$-$'' to $\Gamma_L$ (long decay rate). In the high temperature limit, the decay is exponential with $\Gamma_S=\Gamma_L=(\Gamma_A+\Gamma_F)/2$.

Figures~\ref{fig:s1}b--e show PL dynamics of samples \#1, \#2, \#4, and \#6 at temperatures from 2 to 100 K  measured at emission maxima. The dynamics are multiexponential with an initial fast drop on a timescale of about 3 ns followed by a slower decay. The PL dynamics are quite fast and not sensitive to temperature. We explain it as follows. $\Delta E_{AF}$ is commonly about 1~meV in similar NPLs,\cite{Shornikova2020acsn_s,Smirnova2023_s} which is comparable to $k_B T$ even at cryogenic temperatures. As a result, excitons are distributed between the dark and bright states. The dark state has higher occupation compared to the bright state, but it has slower radiative recombination rate. The resulting emission is contributed by the dark and bright excitons, and the PL dynamics are not sensitive to the changes of the magnetic field and temperature. In other words, the upper bright $\ket{A}$ exciton state is populated at $T=4$ K, and the further increase of the temperature provides no effect to the PL dynamics (compare panels in the bottom of Fig.~\ref{fig:s1}a).

\begin{figure*}[h!]
	\begin{center}
		\includegraphics{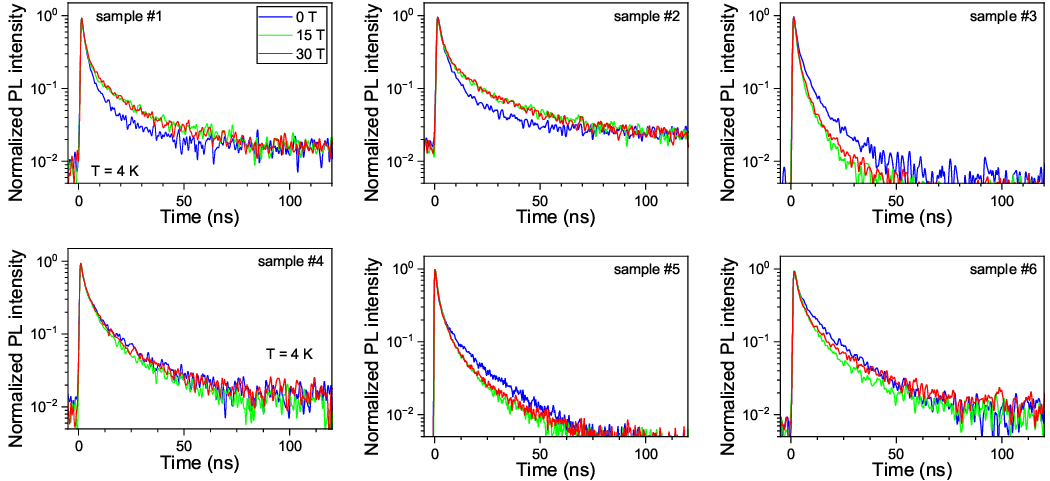}
		\caption{\label{fig:s2}\small Recombination dynamics measured in magnetic fields of 0, 15 and 30~T in samples \#1--\#6 at $T=4$~K.}   
	\end{center}
\end{figure*}

The average exciton decay time is obtained by fitting the PL dynamics with a three-exponential function:
\begin{equation}
	I_{PL}(t) = A_1 e^{-\frac{t}{\tau_1}}+A_2 e^{-\frac{t}{\tau_2}}+A_3 e^{-\frac{t}{\tau_3}},
\end{equation} 
where $\tau_i$ are the decay times of the $i$-th component, and $A_i$ are the corresponding amplitudes. The average decay time can be found as:
\begin{equation}
	\langle \tau \rangle = f_1 \tau_1+f_2 \tau_2+f_3 \tau_3,
\end{equation} 
where $f_i=A_i \tau_i/\sum\limits_{j=1}^3(A_j \tau_j)$ are the corresponding fractions of each component to the total PL intensity.

The average decay times $\langle \tau \rangle$ at $T=4$~K are $15\pm5$~ns in samples \#1 and \#3, $20\pm5$ ns in sample \#2,  $30\pm5$ ns in sample \#4, $20\pm5$ ns in sample \#5, and $40\pm5$ ns in sample \#6.

Figure~\ref{fig:s2} shows the recombination dynamics at various magnetic fields in samples~\#1--\#6 at temperature $T=4$ K. The decays are not affected by magnetic field within our experiment accuracy. As discussed above, it is due to the fact that at $T=4$ K the bright exciton energy level is populated: assuming $\Delta E_{AF}=1$ meV and Boltzmann energy distribution of excitons, the population of the bright state is 5.5\%, which is enough to significantly accelerate the emission decay rate.

The acceleration of the dark exciton radiative decay rate in a magnetic field due to mixing with the bright exciton is described by the following equation:
\begin{equation}
\Gamma_L=\Gamma_F^0+\Gamma_F^B(B,\theta),
\label{eq:MF_rate1}
\end{equation}
where $\Gamma_F^0$ is the radiative recombination rate in zero magnetic field, and $\Gamma_F^B(B,\theta)$ is the radiative rate acquired in a magnetic field tilted with respect
to the quantization axis via admixture of bright exciton states:\cite{Rodina2016_s}
\begin{equation}
	\Gamma_F^B(B,\theta)=\left(\frac{g_e\mu_B B \sin\theta}{2 \Delta E_{AF}}\right)^2 \Gamma_A.
	\label{eq:MF_rate2}
\end{equation}
Here $\Gamma_A$ is the radiative recombination rate of the bright exciton in zero field, $\theta$ is the angle between the magnetic field and NPL quantization axis (axis perpendicular to the NPL plane).

Equations \eqref{eq:MF_rate1} and \eqref{eq:MF_rate2} are applicable in low magnetic fields where $g_e\mu_B B < \Delta E_{AF}$. Since $\Delta E_{AF}\approx1$ meV in the studied NPLs, this condition is fulfilled for magnetic fields below about 10~T. In stronger magnetic fields, the dark exciton radiative decay further accelerates, and the bright exciton radiative rate decelerates, so that in high magnetic fields the decay is exponential with $\Gamma_L=\left(\Gamma_A+\Gamma_F^0\right)/2$. To simplify the modeling, we assume that the decay time  $\langle \tau \rangle$ is the same in all our samples and equals to 20 ns. The average decay rate $\Gamma_L = 1/\langle \tau \rangle=0.05$ ns$^{-1}$. This allow us to estimate $\Gamma_A=2\Gamma_L=0.1$~ns$^{-1}$.

We neglected the acceleration of the decay in magnetic fields in our modeling (see Section S5), since the main effect on nonlinear DCP with sign reverse comes from  thermal population of the exciton states. We use $\Gamma_F=0.01$~ns$^{-1}$ for our modeling.

\vspace{10mm}

\textbf{\large S2. Additional experimental data for CdSe/CdZnS and CdSe/CdZnMnS NPLs}

\begin{figure*}[h!]
	\begin{center}
		\includegraphics{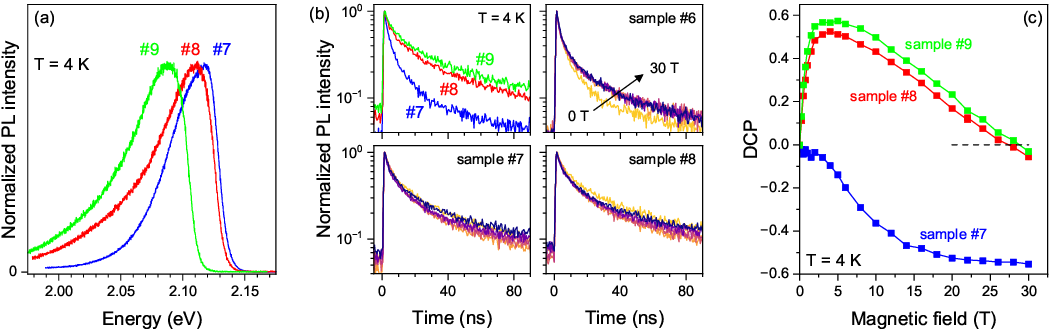}
		\caption{\label{fig:s3}\small Emission properties and degree of circular polarization (DCP) of CdSe/CdZnS and CdSe/CdZnMnS NPLs. Blue: CdSe/CdZnS NPLs (sample \#7), red and green: CdSe/CdZnMnS NPLs with 4 ML (sample \#8) and 5ML (sample \#9) thick shells, respectively. (a) Normalized PL spectra at $T=4$ K. (b) Top left: Emission dynamics in PL maxima at $T=4$ K in Samples \#7--\#9. Other panels: PL dynamics at various magnetic fields from $B=0$ (yellow) to $30$~T (purple) with a step of 6~T.  (c) DCP as a function of the magnetic field.}   
	\end{center}
\end{figure*}

We measured additionally CdSe/CdZnS and CdSe/CdZnMnS NPLs, the parameters of the samples \#7--\#9 are listed in Table~\ref{tabs1}. The synthesis details are provided in Methods. Figure~\ref{fig:s3}a shows the PL spectra of these NPLs. Samples \#7 and \#8 have similar structures (2 ML core and 4 ML shells), but a small amount of Mn is added to the shell of the sample \#8. Their emission spectra are centered at about the same energy of 2.118 eV (\#7) and 2.110 eV (\#8). Sample \#9 has the same 2 ML cores but thicker 5 ML DMS shells, therefore, its emission is slightly shifted to lower energies having maximum at 2.090 eV. Recombination dynamics measured in the emission maxima are shown in Fig.~\ref{fig:s3}b. Sample \#7 has a slightly faster PL decay, which we cannot explain. Similarly to samples \#1--\#6, the emission decays are not sensitive to magnetic field, because the bright exciton state is populated at $T=4$~K.

\begin{table}
	\small
	\caption{Parameters of the studied NPL samples \#7--\#9.}
	\begin{tabular}{ |c|c|c|c| } 
		\hline
		Sample & Core thickness & Shell thickness & Total thickness\\ 
		\hline
		\#7 & CdSe, 2 ML & CdZnS, 4 ML & 10 ML\\
		\#8 & CdSe, 2 ML & CdZnMnS, 4 ML & 10 ML\\ 
		\#9 & CdSe, 2 ML & CdZnMnS, 5 ML & 12 ML\\ 
		\hline
	\end{tabular}
	\label{tabs1}
\end{table}

Figure~\ref{fig:s3}c shows magnetic field dependence of DCP in samples \#7--\#9 measured at $T=4$ K. Generally, these samples behave similar to samples \#1--\#6. In the nonmagnetic sample \#7, the DCP is negative. Its absolute value monotonously increases with magnetic field strength, and starts to saturate at $B>20$ T reaching $-0.55$ at $B=30$ T. In the DMS samples \#8 and \#9 the DCP is positive in low magnetic fields reaching +0.52 and +0.57 in samples \#8 and \#9, respectively, in $B=4$ T. In higher magnetic fields the DCP decreases, crosses zero at $B_c=27$ and 29 T in samples \#8 and \#9, respectively, and continues to negative values for $B>B_c$.

%\vspace{15mm}
\clearpage
\textbf{\large S3. Band structure of CdSe/CdMnS NPLs. Electron and hole penetration into shells}

Here we reproduce the model developed in Refs.\onlinecite{Shornikova2020acsn_s, Delikanli2015_s} and calculate electron and hole penetration into shells. We use a particle in a box model and assume that: (i) there is no confinement in the nanoplatelet plane; (ii)
the electron-hole Coulomb interaction is negligible; (iii) the effective mass approximation is valid; and (iv) infinite potential barriers exist at the external surfaces.

\begin{figure*}[h!]
	\begin{center}
		\includegraphics{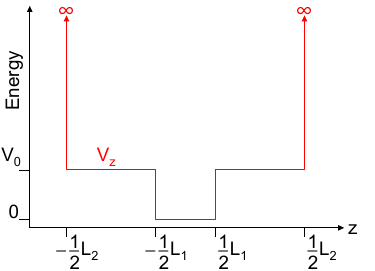}
		\caption{\label{fig:s4} Energy diagram for the particle in a box model. $L_1$ is the thickness of the core, $L_2$ is the total thickness of NPL, $V_z$ is the potential energy.}   
	\end{center}
\end{figure*}

The stationary Schrödinger equation is given by
\begin{equation}
	\left[-\frac{\hbar^2}{2m}\frac{d^2}{dz^2}+V_z(z)\right]\Psi(z)=E_z\Psi(z).
	\label{eq:Schroedinger}
\end{equation}
Here $m$ is the effective mass, and $V_z(z)$ is the potential energy with a shape shown in Fig.~\ref{fig:s4}. For electrons $V_0=\Delta E_c$, and for holes $V_0=\Delta E_v$, where $\Delta E_c$ and $\Delta E_v$ are the conduction and valence band offsets, respectively.

The boundary conditions are:

\begin{equation}
	\begin{cases}
		\Psi_1(-\frac{1}{2}L_2)=0\\
		\Psi_1(-\frac{1}{2}L_1)=\Psi_2(-\frac{1}{2}L_1)\\
		\Psi_2(\frac{1}{2}L_1)=\Psi_3(\frac{1}{2}L_1)\\
		\Psi_3(\frac{1}{2}L_2)=0\\
		\frac{1}{m_1}\Psi^\prime_1(-\frac{1}{2}L_1)=\frac{1}{m_2}\Psi^\prime_2(-\frac{1}{2}L_1)\\
		\frac{1}{m_2}\Psi^\prime_2(\frac{1}{2}L_1)=\frac{1}{m_1}\Psi^\prime_3(\frac{1}{2}L_1).
	\end{cases}       
\end{equation}
Here $\Psi_1\in[-\frac{1}{2}L_2;-\frac{1}{2}L_1]$, $\Psi_2\in[-\frac{1}{2}L_1;\frac{1}{2}L_1]$, and $\Psi_3\in[\frac{1}{2}L_1;\frac{1}{2}L_2]$.

The normalization condition is given by: $\int\limits_{-\frac{1}{2}L_2}^{-\frac{1}{2}L_1}\Psi^2_1(z)dz+\int\limits_{-\frac{1}{2}L_1}^{\frac{1}{2}L_1}\Psi^2_2(z)dz+\int\limits_{\frac{1}{2}L_1}^{\frac{1}{2}L_2}\Psi^2_3(z)dz=1$.

The Schrödinger equation can be rewritten as:

%\begin{equation}
%	\begin{cases}
%		\Psi_1^{\prime\prime}+\frac{2m_1}{\hbar^2}(E-V_0)\Psi_1=0 & -\frac{1}{2}L_2<z<-\frac{1}{2}L_1\\
%		\Psi_2^{\prime\prime}+\frac{2m_2}{\hbar^2}E\Psi_2=0 & -\frac{1}{2}L_1<z<\frac{1}{2}L_1\\
%		\Psi_3^{\prime\prime}+\frac{2m_1}{\hbar^2}(E-V_0)\Psi_3=0 & \frac{1}{2}L_1<z<-\frac{1}{2}L_2
%	\end{cases}
%	\label{eq:Schroedinger2}      
%\end{equation}

\begin{equation}
	\begin{cases}
		\Psi_1^{\prime\prime}+\frac{2m_1}{\hbar^2}(E-V_0)\Psi_1=0\\
		\Psi_2^{\prime\prime}+\frac{2m_2}{\hbar^2}E\Psi_2=0\\
		\Psi_3^{\prime\prime}+\frac{2m_1}{\hbar^2}(E-V_0)\Psi_3=0.
	\end{cases}
	\label{eq:Schroedinger2}      
\end{equation}

The solutions of eq~\eqref{eq:Schroedinger2} are well known. There are two possible cases.

\vspace{5mm}
\textbf{I.} For $E<V_0$ the Schrödinger equation takes the following form:
\begin{equation}
	\begin{cases}
		\Psi_1^{\prime\prime}-k_1^2\Psi_1=0\\
		\Psi_2^{\prime\prime}+k_2^2\Psi_2=0\\
		\Psi_3^{\prime\prime}-k_1^2\Psi_3=0,
	\end{cases}
	\label{eq:Schroedinger3}      
\end{equation}
where $k_1=\sqrt{\frac{2m_1}{\hbar^2}(V_0-E)}$, $k_2=\sqrt{\frac{2m_2}{\hbar^2}E}$, $m_1$ and $m_2$ are the effective masses in the shell and in the core, respectively, $L_1$ is the thickness of the core, $L_2$ is the total thickness of NPL. The wave function is of the form

\begin{equation}
	\Psi(z) =
	\begin{cases}
		\Psi_1(z)=A_1e^{k_1 z} + B_1e^{-k_1 z} \\
		\Psi_2(z)=A_2\cos(k_2 z) + B_2\sin(k_2 z) \\
		\Psi_3(z)=A_3e^{k_1 z} + B_3e^{-k_1 z}.
	\end{cases}       
\end{equation}

For even wave functions (including the lowest energy), the energies are solutions of the following equation:
\begin{equation}
	\frac{k_1}{m_1}\coth\frac{k_1(L_2-L_1)}{2}=\frac{k_2}{m_2}\tan\frac{k_2 L_1}{2}.
\end{equation}

For the odd wave functions, the equation reads as:

\begin{equation}
	\frac{k_1}{m_1}\coth\frac{k_1(L_2-L_1)}{2}=-\frac{k_2}{m_2}\cot\frac{k_2 L_1}{2}.
\end{equation}

\textbf{II.} For $E>V_0$ the Schrödinger equation takes the following form:
\begin{equation}
	\begin{cases}
		\Psi_1^{\prime\prime}+k_1^2\Psi_1=0\\
		\Psi_2^{\prime\prime}+k_2^2\Psi_2=0\\
		\Psi_3^{\prime\prime}+k_1^2\Psi_3=0,
	\end{cases}
	\label{eq:Schroedinger3}      
\end{equation}
where $k_1=\sqrt{\frac{2m_1}{\hbar^2}(E-V_0)}$. The wave function is of the form
\begin{equation}
	\Psi(z) =
	\begin{cases}
		\Psi_1(z)=A_1\cos(k_1 z) + B_1\sin(k_1 z) \\
		\Psi_2(z)=A_2\cos(k_2 z) + B_2\sin(k_2 z) \\
		\Psi_3(z)=A_3\cos(k_1 z) + B_3\sin(k_1 z).
	\end{cases}       
\end{equation}

For even wave functions (including the lowest energy), the energies are solutions of the following equation:
\begin{equation}
	\frac{k_1}{m_1}\cot\frac{k_1(L_2-L_1)}{2}=\frac{k_2}{m_2}\tan\frac{k_2 L_1}{2}.
\end{equation}
For the odd wave functions, the equation reads as:
\begin{equation}
	\frac{k_1}{m_1}\cot\frac{k_1(L_2-L_1)}{2}=-\frac{k_2}{m_2}\cot\frac{k_2 L_1}{2}.
\end{equation}

We calculated the energies and carrier wave functions using the following parameters. One ML thickness equals to $0.35$~nm.\cite{Mahler2012_s} To account for the fact that the NPLs are Cd-terminated, 0.5 ML was added to the total thickness. Electron effective masses are $m_e^{\rm CdSe}=0.18m_0$ and $m_e^{\rm CdS}=0.35m_0$, hole effective masses are $m_h^{\rm CdSe}=0.89m_0$ and $m_h^{\rm CdS}=0.95m_0$, where $m_0$ is the free electron mass.\cite{Ithurria2011_s,Strassberg2019_s,Muckel2018_s}

Important parameters of the calculation are the band offsets. The low temperature bandgap values are $E_g^{\rm CdSe}\approx1.75$~eV and $E_g^{\rm CdS}\approx2.5$ eV.\cite{Adachi2004_s,Adachi2005_s} This value is valid for CdSe NPLs with zinc blend crystal structure. Note that $E_g^{\rm CdSe}=1.84$ eV used in Refs.\onlinecite{Strassberg2019_s,Muckel2018_s} corresponds to wurtzite CdSe. The bandgap difference between the CdSe core and CdS shell equals to 0.75 eV.

The conduction and valence band offsets between CdSe and CdS are not precisely known. However, the valence band offsets $\Delta E_v$ reported in the literature are large, at least 0.45~eV, so that the hole is believed to be well confined in the CdSe core. The reported conduction band offsets $\Delta E_c$ range from $0.3$~eV to 0~eV, and this value depends on the crystal structure, NC size, lattice strain and temperature, see Ref.\onlinecite{Javaux2013_s} for references. Due to weak, if present at all, confinement, the electron wave function penetrates into the CdS shell, while the hole wave function is localized in the core. For this reason, CdSe/CdS nanostructures are usually called quasi-type II.

The calculated probabilities to find an electron and a hole in the shells, $f_e$ and $f_h$, respectively, are listed in Table~\ref{tabs2} for two limiting cases: $\Delta E_c=0.3$ eV and $\Delta E_v=0.45$ eV, and $\Delta E_c=0$ eV and $\Delta E_v=0.75$ eV. The first case was used for the calculations of exciton levels and DCP of emission in magnetic fields (Table~\ref{tab1}).

\begin{table}
	\small
	\caption{Probabilities to find an electron and a hole in the shells}
	\begin{tabular}{ |c|c|c|c|c|c|c|} 
		\hline
		{Sample} &$L_1$, ML&$L_2$, ML& \multicolumn{2}{c|}{$f_e$}&\multicolumn{2}{c|}{$f_h$}\\
		\cline{4-7}
		&&& $\Delta E_c=0.3$ eV & $\Delta E_c=0$ & $\Delta E_v=0.45$ eV&$\Delta E_v=0.75$ eV\\
		\hline
		\#1 &2&6&  0.37  &  0.41  &  0.22  &  0.16\\
		\#2 &2&6&  0.37  &  0.41  &  0.22  &  0.16\\ 
		\#3 &2&14&  0.58  &  0.72  &  0.26  &  0.17\\
		\#4 &4&6&  0.08  &  0.09  &  0.03  &  0.03\\
		\#5 &4&8&  0.18  &  0.22  &  0.06  &  0.04\\ 
		\#6 &4&10&  0.24  &  0.33  &  0.07  &  0.05\\ 
		\hline
	\end{tabular}
	\label{tabs2}
\end{table}

\vspace{15mm}
%\clearpage
\textbf{\large S4. Exchange interaction of charge carriers with Mn$^{2+}$ ions}
%\vspace{3mm}

Here we reproduce the approach developed in Ref.\onlinecite{Shornikova2020acsn_s}, see Supporting Information S6 of Ref.\onlinecite{Shornikova2020acsn_s}.

To model the exchange interaction of charge carriers with Mn$^{2+}$ ions, we follow the conventional approach for diluted magnetic semiconductors based on consideration of the exchange interaction within the mean field approximation. The exchange interaction is commonly explained in terms of the giant Zeeman splitting. The Zeeman splitting of a particle (exciton, charge carrier, etc.) is expressed as a sum of two parts:
\begin{equation}
	\Delta E_Z(B)=\Delta E_{Z}^{\rm intr}(B)+\Delta E_{Z}^{\rm exch}(B).
\end{equation}
Here $\Delta E_{Z}^{\rm intr}(B)$ is the intrinsic Zeeman splitting, which is present in nonmagnetic NPLs. In a magnetic field $\vec{\boldsymbol{B}}$, an electron with a magnetic moment $\vec{\boldsymbol{\mu}}_e$ acquires an additional energy $\Delta E=-\vec{\boldsymbol{\mu}}_e\cdot\vec{\boldsymbol{B}}$. The electron magnetic moment can be written as: $\vec{\boldsymbol{\mu}}_e=-\mu_B g_e \vec{\boldsymbol{J}}/\hbar$, where $\mu_B$ is the Bohr magneton, $g_e$ is the Landé $g$-factor of electron, and $\vec{\boldsymbol{J}}$ is the total electronic angular momentum. Electrons with the angular momentum projections onto the quantization axis $J_z=+1/2$ and $J_z=-1/2$ acquire different energies in a magnetic field, and the splitting between the electron levels with angular momentum projections $\ket{\pm 1/2}$ appears. For electrons this splitting equals to:
\begin{equation}
	\Delta E_{Z,e}^{\rm intr}(B)=g_e \mu_B B,
\end{equation} 
and for heavy holes with $J_z=\pm 3/2$:
\begin{equation}
	\Delta E_{Z,h}^{\rm intr}(B)=-3g_h \mu_B B,
	\label{eq:s10}
\end{equation} 
where $g_h$ is the hole $g$-factor. Here we assume that the heavy hole is the ground hole state, which is valid for CdSe-based NPLs.\cite{Shornikova2018_s,Biadala2014_s}

$\Delta E_{Z}^{\rm exch}(B)$ represents the exchange energy due to the interaction of carriers with the Mn$^{2+}$ ions. This energy for electrons equals to:
\begin{equation}
	\Delta E_{Z,e}^{\rm exch}=-\langle S_{\rm Mn} \rangle x N_0 \alpha f_e,
\end{equation}
and for holes to:
\begin{equation}
	\Delta E_{Z,h}^{\rm exch}=-\langle S_{\rm Mn} \rangle x N_0 \beta f_h.
\end{equation}
Here $x$ is the Mn concentration, $N_0$ is the number of cations per unit volume, $\alpha$ and $\beta$ are the \textit{s}--\textit{d} and \textit{p}--\textit{d} exchange constants in Cd$_{1-x}$Mn$_x$S, respectively, $f_e$ and $f_h$ are the probabilities to find an electron and a hole in the shells, respectively, and $\langle S_{\rm Mn} \rangle$ is the mean spin of Mn$^{2+}$ ions. $\langle S_{\rm Mn} \rangle$ depends on  on the external magnetic field, the temperature and the Mn concentration: $\langle S_{\rm Mn}\rangle=f(x,T,B)$. The latter dependence is provided by the fact  that neighboring Mn$^{2+}$ ions interact antiferromagnetically with each other, which reduces $\langle S_{\rm Mn}\rangle$ with increasing $x$.  To account for this, we use the function for the mean spin from Refs.\onlinecite{Kossut2010_s,Keller2001_s}:
\begin{equation}
	\langle S_{\rm Mn}\rangle=-S_0(x)B_S\left(\frac{g_{\rm Mn}\mu_B SB}{k_B(T+T_0(x))}\right).
	\label{eq:s13}
\end{equation}
Here $B_S (\xi)$ is the Brillouin function, $B_S(\xi)=\frac{2S+1}{2S}\coth\left(\frac{2S+1}{2S}\xi\right)-\frac{1}{2S}\coth\left(\frac{1}{2S}\xi\right)$. $S=5/2$ is the spin of Mn$^{2+}$, $g_{\rm Mn}=2.01$ is the $g$-factor of Mn$^{2+}$, $S_0(x)=-0.804+0.364/(x+0.109)$ is the effective spin, and $T_0(x)=47.2x-281x^2+714x^3$ is the effective temperature. Note that these $S_0(x)$ and $T_0(x)$ functions have been reported for Zn$_{1-x}$Mn$_x$Se.\cite{Keller2001_s} The parameters $S_0(x)$ and $T_0(x)$ describe phenomenologically the interaction between the Mn$^{2+}$ ions. The Mn--Mn interaction is modified in thin layers compared to bulk, because Mn$^{2+}$ ions have less neighbors for same $x$.\cite{Yakovlev1995,Kneip2006} In our study, Mn concentrations are small and such deviations are negligible. %\cD{For CdMnTe there is data collection in Fig. 1.3 of Dima Doctor thesis on magnetic polaron. But I can not find interpolation parameters. I would expect that they are not far from ZnMnSe. Would be interesting to check. }

\begin{figure*}[h!]
	\begin{center}
		\includegraphics{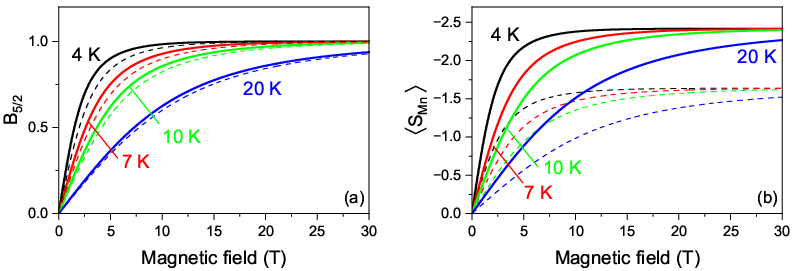}
		\caption{\label{fig:s5} (a) Brillouin function $B_{5/2}$ at various temperatures for $x=0.004$ (solid lines) and $x=0.04$ (dashed lines). The Brillouin function weakly depends on Mn concentration. (b) Mean spin of Mn$^{2+}$ ions at various temperatures for $x=0.004$ (solid lines) and $x=0.04$ (dashed lines). The amplitude of the $\langle S_{\rm Mn}\rangle$ function depends on the Mn concentration, while the shape is the Brillouin function $B_{5/2}$.}   
	\end{center}
\end{figure*}

Figure~\ref{fig:s5}a shows the Brillouin function $B_{5/2}(\xi)$ at various temperatures for two Mn concentrations of $x=0.004$ (solid lines) and $x=0.04$ (dashed lines). The shape of the Brillouin function weakly depends on Mn concentration. Figure~\ref{fig:s5}b shows the mean Mn spin at various temperatures for two Mn concentrations of $x=0.004$ (solid lines) and $x=0.04$ (dashed lines). The shape of the $\langle S_{\rm Mn}\rangle$ function repeats the shape of the $B_{5/2}(\xi)$ function and weakly depends on the Mn concentration, while its amplitude equals to $S_0(x)$ and strongly varies with the Mn concentration.

\begin{figure*}[h!]
	\begin{center}
		\includegraphics{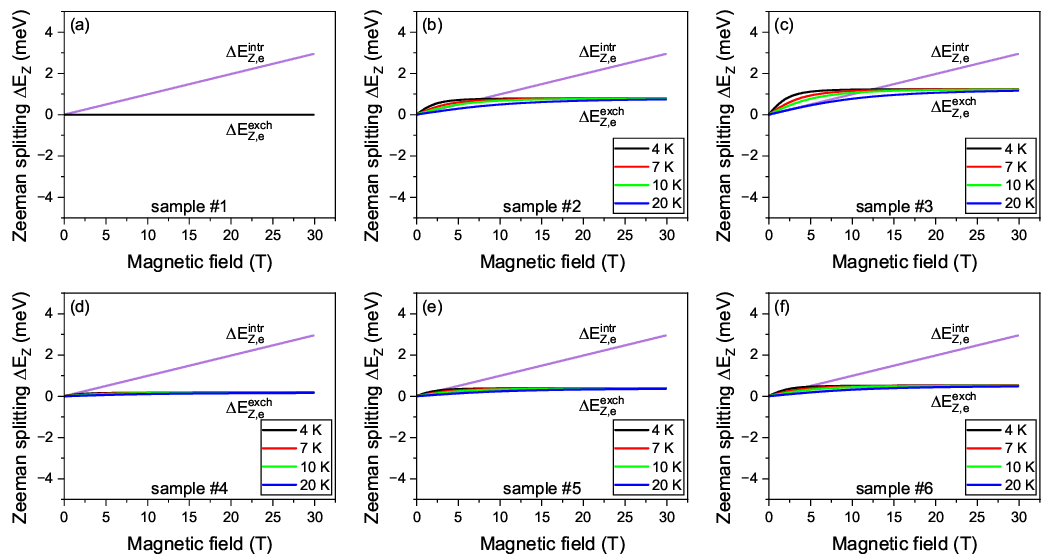}
		\caption{\label{fig:Sintr-exch_e} \small Calculated exchange and intrinsic Zeeman splitting of electrons as function of the magnetic field at various temperatures in samples \#1--\#6 with parameters as in Fig.~\ref{fig4}.}   
	\end{center}
\end{figure*}

\begin{figure*}[h!]
	\begin{center}
		\includegraphics{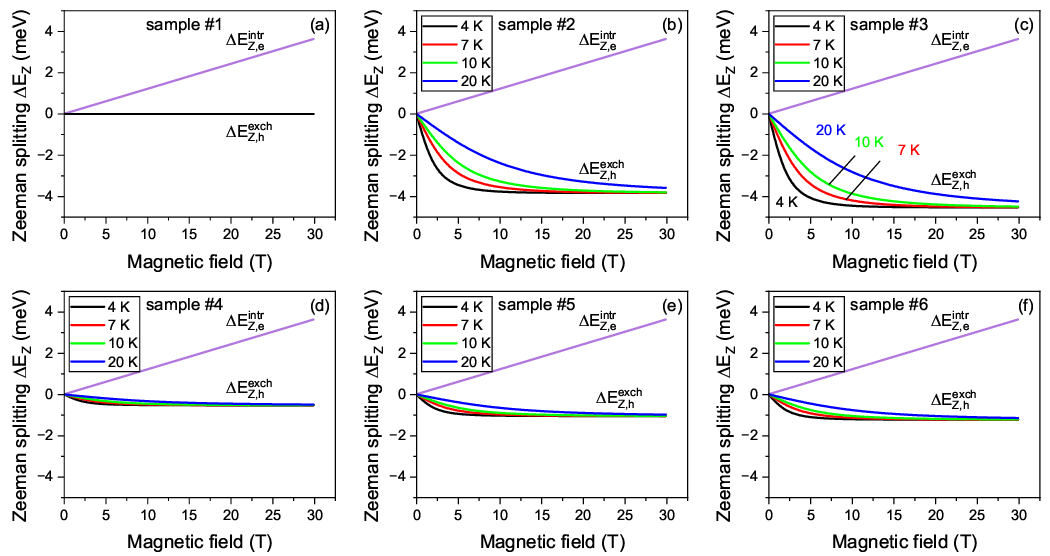}
		\caption{\label{fig:Sintr-exch_h}\small Calculated exchange and intrinsic Zeeman splitting of holes as function of the magnetic field at various temperatures in samples \#1--\#6 with parameters as in Fig.~\ref{fig4}.}   
	\end{center}
\end{figure*} 

Figures~\ref{fig:Sintr-exch_e} and \ref{fig:Sintr-exch_h} show intrinsic and exchange Zeeman splittings of  electrons and holes in magnetic fields at various temperatures. The following parameters were used for the calculation: $g_e=1.7$ (determined from spin-flip Raman scattering),\cite{Shornikova2020acsn_s} $g_h=-0.7$,\cite{Shornikova2020acsn_s,Shornikova2018nl_s} $N_0\alpha=0.22$~eV and $N_0\beta=-1.8$~eV \cite{Kossut2010_s}, $S_0(x=0.004)=2.42$ and $T_0(x=0.004)=0.18$~K. Note that in Ref.\onlinecite{Shornikova2018nl_s} the hole $g$-factor increased with the magnetic field strength from $g_h=-0.4$ in $B=0$ to $g_h=-0.7$ in $B=15$~T due to band mixing effects. Here we use the high magnetic field value.

The intrinsic splittings increase linearly with the magnetic field strength, while the exchange energy rapidly increases in low magnetic fields, and then saturates. Since the average Mn spin depends on temperature (Fig.~\ref{fig:s5}), the exchange splitting also varies with temperature.

\begin{figure*}[t!]
	\begin{center}
		\includegraphics{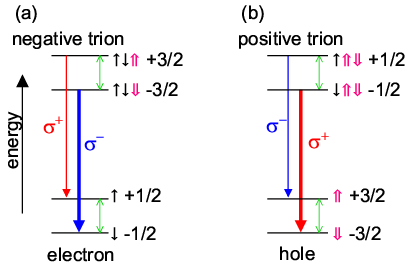}
		\caption{\label{fig:Fig_SI_T_schemes}\small Schematics of the spin level structure and optical transitions for negatively and positively charged trions in a magnetic field for CdSe-based NPLs with $g_e>0$ and $g_h<0$.~\cite{Shornikova2018nl_s} The electron and hole Zeeman splittings are shown by the thin green arrows. The short black and pink arrows indicate the electron and hole spins, respectively, and the numbers give the trion spin projections. The polarized optical transitions are shown by the red ($\sigma^+$) and blue ($\sigma^-$) arrows. The more intense emission, shown by the thicker arrows, comes from the lowest-energy trion state with spin $-3/2$ for the negative trion, and with spin $-1/2$ for the positive trion.}   
	\end{center}
\end{figure*}

Trions are charged excitons, which have one additional charge carrier, electron or hole. A negative trion consists of two electrons in a singlet state and one unpaired hole, so that the electron spins compensate each other and the hole Zeeman splitting determines the trion splitting. Similarly, a positive trion consists of two holes in a singlet state and one unpaired electron, so that the hole spins compensate each other and the electron Zeeman splitting  determines the trion splitting (Fig.~\ref{fig:Fig_SI_T_schemes}). This splitting also determines the sign of the degree of circular polarization (DCP).

The DCP is given by
\begin{equation} \label{eq:s23-DCP}
	P_c(B)= P^{sat} \tanh  \frac{\Delta E_Z(B)}{2k_B T} \, ,
\end{equation} 
where $P^{sat}$ is the saturation degree, which depends on the particle spin structure and NPL orientation in the ensemble. Note that the sign of $P^{sat}$ depends on the origin of the emitting particle. In CdSe-based nonmagnetic NPLs, the PL is $\sigma^-$ polarized for bright and dark excitons.

The sign of the polarization of trions is determined by the charge (positive or negative) and the carrier $g$-factors. In CdSe-based nonmagnetic NCs $g_e>0$ and $g_h<0$, and the intrinsic DCP is $\sigma^+$ polarized for positive trions, because the $\ket{-1/2}$ electron level has lower energy than the $\ket{+1/2}$ level. The intrinsic DCP is $\sigma^-$ polarized for negative trions, because the $\ket{-3/2}$ hole level has lower energy than the $\ket{+3/2}$ level.

\begin{figure*}[t!]
	\begin{center}
		\includegraphics{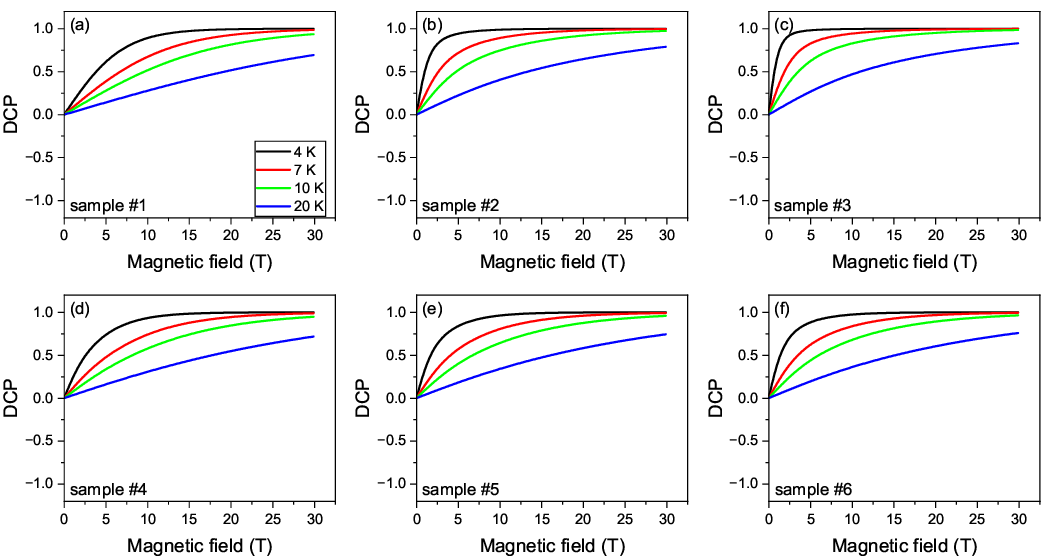}
		\caption{\label{fig:Fig_DCP_e_Tpos}\small Calculated DCP of positive trion using the trion energy levels shown in Fig.~\ref{fig:Sintr-exch_e} as functions of the magnetic field at various temperatures in samples \#1--\#6 with parameters as in Fig.~\ref{fig4}.}   
	\end{center}
\end{figure*} 

\begin{figure*}[t!]
	\begin{center}
		\includegraphics{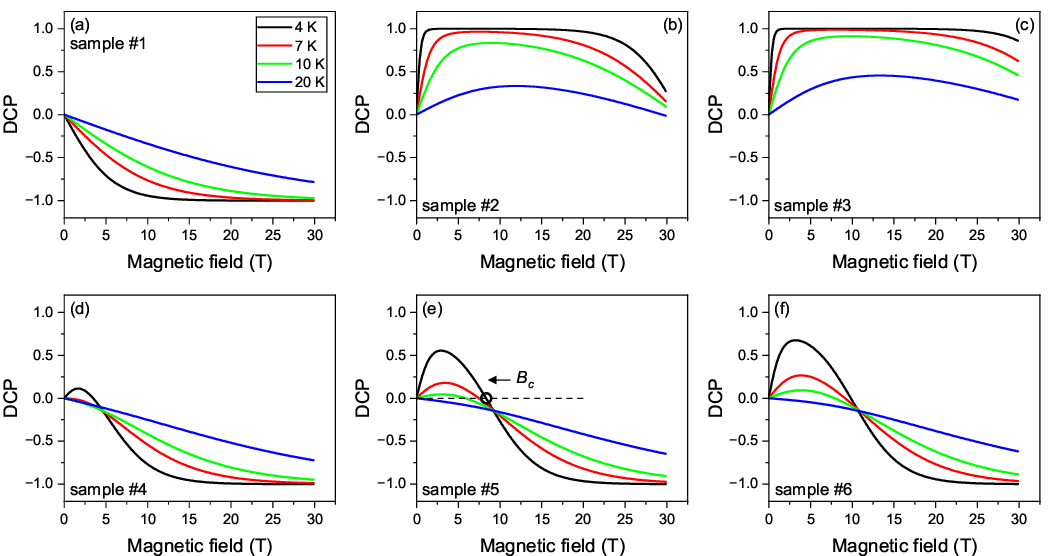}
		\caption{\label{fig:Fig_DCP_h_Tneg}\small Calculated DCP of negative trion using the trion energy levels shown in Fig.~\ref{fig:Sintr-exch_h} as functions of the magnetic field at various temperatures in samples \#1--\#6 with  parameters as in Fig.~\ref{fig4}.}   
	\end{center}
\end{figure*} 

In DMS samples, the interaction of electrons in positive trions and holes in negative trions with Mn$^{2+}$ spins, brings the additional energy $\Delta E_Z^{\rm exch}$. This energy is small for electrons and has the same sign as the intrinsic Zeeman splitting (Fig.~\ref{fig:Sintr-exch_e}). Therefore, it is expected that the DCP of positive trions has $\sigma^+$ polarization and increases monotonously with magnetic field in nonmagnetic and DMS NPLs (Fig.~\ref{fig:Fig_DCP_e_Tpos}). This behavior is different from the experimentally measured data. We conclude from that that the positive trions are absent in the emission of the studied NPLs, or their impact is very small and can be neglected.

The intrinsic and exchange Zeeman splittings of holes are comparable and have opposite signs (Fig.~\ref{fig:Sintr-exch_h}). The DCP calculated as functions of magnetic field at various temperatures for negative trions are shown in Fig.~\ref{fig:Fig_DCP_h_Tneg}. The DCP of negative trions in nonmagnetic samples is $\sigma^-$ polarized. In DMS samples, the DCP is $\sigma^+$ polarized in low magnetic fields and switches the sign in high magnetic fields. When the temperature increases, the intrinsic Zeeman splitting does not change, while the exchange splitting reduces (Figs.~\ref{fig:Sintr-exch_h}b--f). Due to this, the critical magnetic field $B_c$ is smaller at higher temperatures (Figs.~\ref{fig:Fig_DCP_h_Tneg}b--f). This is opposite to the experimentally observed increase of $B_c$ with growing temperature, see Figs.~\ref{fig2}d~\ref{fig3}d,e. We conclude from this that the negative trions are absent in the emission of the samples, or their impact is very small and can be neglected.

%Note that negative trions have nonlinear DCP only if their intrinsic Zeeman splitting is comparable with the exchange splitting. The intrinsic Zeeman splitting is controlled by the hole $g$-factor $g_h$ (eq.~\eqref{eq:s10}). We used the high-field hole $g$-factor value $g_h=-0.7$\cite{Shornikova2018nl_s} in our calculations. The theoretical calculations predict $g_h=-0.2$.\cite{Shornikova2018nl_s}  With $g_h=-0.2$, the intrinsic Zeeman splitting is not strong enough to compensate the exchange energy in magnetic fields up to 30 T in samples \#2--\#6 (not shown). 

\vspace{15mm}
%\clearpage
\textbf{\large S5. Exchange interaction of excitons with Mn$^{2+}$ ions}
%\vspace{3mm}

Following the same approach, the exciton Zeeman splitting and exchange energy can be written as a sum of the electron and hole parts. For the bright exciton, the intrinsic Zeeman splitting reads as
\begin{equation} \label{e3-1}
	\Delta E_{\rm Z,XA}^{\rm intr}(B)= g_{\rm XA}\mu_B B=(-g_e-3g_h)\mu_B B\, ,
\end{equation} 
where $g_{\rm XA}=-g_e-3g_h$ is the bright exciton $g$-factor. In DMS samples an additional term, $\Delta E_{Z,XA}^{\rm exch}(B)$, describing the exciton exchange interaction with the Mn$^{2+}$ spins is added:

\begin{figure*}[h!]
	\begin{center}
		\includegraphics{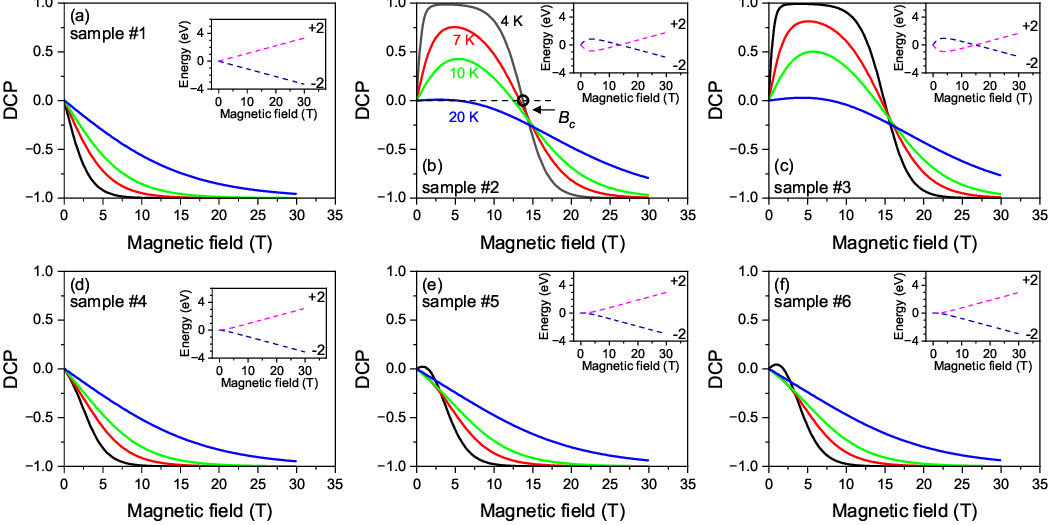}
		\caption{\label{fig:Fig_4_DE}\small Calculated DCP of the dark exciton as a function of magnetic field at various temperatures in samples \#1--\#6 with  parameters  as in Fig.~\ref{fig4}. This corresponds to $\Delta E_{AF}=\infty$. Insets show calculated exciton energy levels at $T=4$ K.}   
	\end{center}
\end{figure*}

\begin{equation} \label{e3-2}
	\Delta E_{\rm Z,XA}(B)= g_{\rm XA}\mu_B B+ \Delta E_{Z,XA}^{\rm exch}(B)=(-g_e-3g_h)\mu_B B+\langle S_{\rm Mn}\rangle x\left(N_0\alpha f_e - N_0\beta f_h\right)\, .
\end{equation}
Note that $\Delta E_{Z,XA}^{\rm exch}(B)$ is controlled by the exchange interaction with the Mn$^{2+}$ ions of both electron and hole composing the exciton.

\begin{figure*}[t!]
	\begin{center}
		\includegraphics{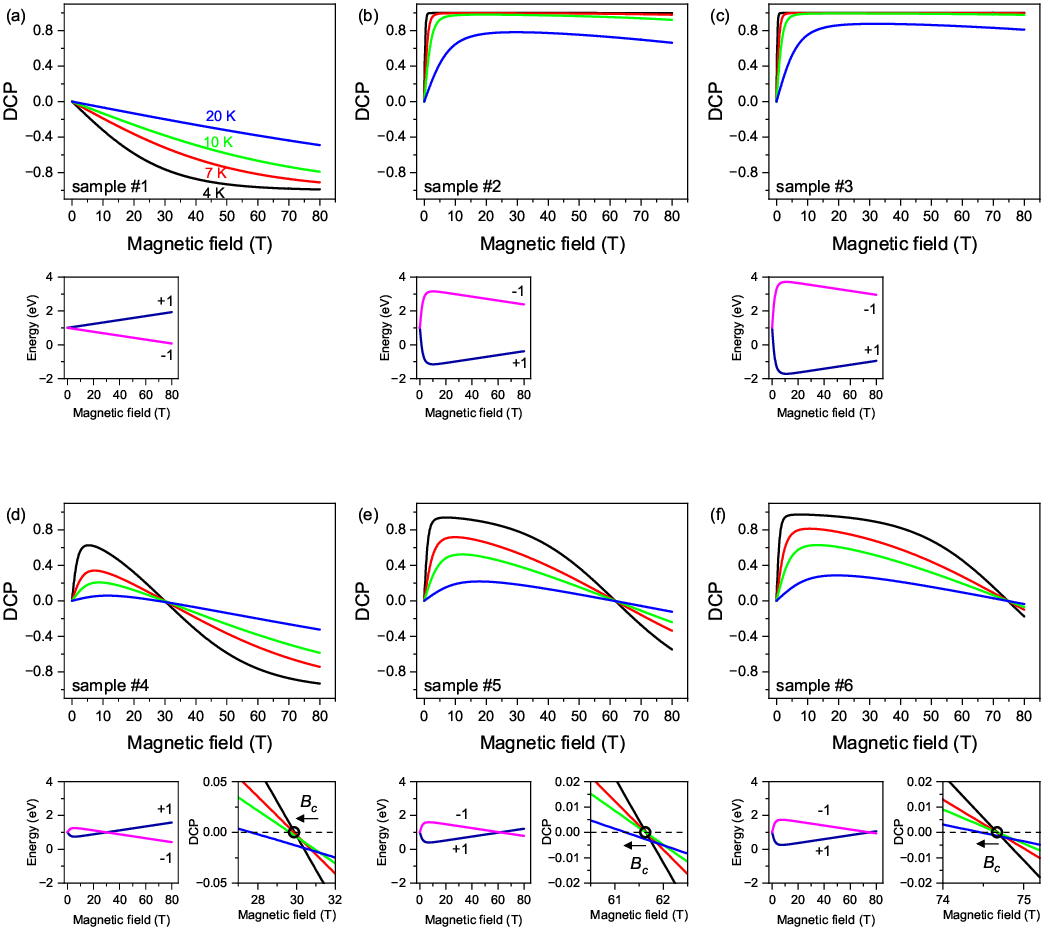}
		\caption{\label{fig:Fig_4_BE}\small Calculated DCP of the bright exciton as a function of magnetic field at various temperatures in samples \#1--\#6 with parameters as in Fig.~\ref{fig4}. Additional panels below the main plots show calculated exciton energy levels at $T=4$ K and the region where DCP crosses zero in a greater detail. Note that horizontal scale is different than in Figs.~\ref{fig:Sintr-exch_e}--\ref{fig:Fig_4_DE}.}   
	\end{center}
\end{figure*}

The intrinsic dark exciton Zeeman splitting is
\begin{equation} \label{e3-3}
	\Delta E_{\rm Z,XF}(B)= g_{\rm XF}\mu_B B=(g_e-3g_h)\mu_B B\, ,
\end{equation} 
where $g_{\rm XF}=g_e-3g_h$ is the dark exciton $g$-factor. In DMS samples, the Zeeman splitting reads as 
\begin{equation} \label{eq:s_DE}
	\Delta E_{\rm Z,XF}(B)= g_{\rm XF}\mu_B B+ \Delta E_{\rm XF}^{\rm exch}(B)=(g_e-3g_h)\mu_B B+\langle S_{\rm Mn}\rangle x\left(-N_0\alpha f_e - N_0\beta f_h\right)\, .
\end{equation}

Insets in Fig.~\ref{fig4} in the main text show the Zeeman splittings of the bright and dark excitons  as functions of the magnetic field. The parameters used for the calculation are the same as in Section~S4: $g_e=1.7$, $g_h=-0.7$, $N_0\alpha=0.22$~eV, $N_0\beta=-1.8$~eV, $x=0.004$, $S_0(x)=2.42$, and $T_0(x)=0.18$~K.  The dark-bright exciton splitting value is taken as $\Delta E_{AF}=1$~meV.  

The exciton DCP is given by eq~\eqref{eq:s23-DCP}. Figure~\ref{fig4} in the main text shows DCP \textit{vs} magnetic field for samples \#1--\#6 at various temperatures. The DCP is calculated for $|P^{sat}|=1$, assuming a Boltzmann distribution between the exciton levels.

We would like to point out that the competition between the intrinsic and exchange Zeeman splitting in dark exciton (without interplay with bright exciton) can only provide a nonlinear DCP, but does not explain the shift of the critical magnetic field $B_c$ to higher fields with increasing temperature. Figure~\ref{fig:Fig_4_DE} shows the DCP and energy levels (insets) for the dark excitons calculated assuming that $\Delta E_{AF}=\infty$ and other parameters as in Fig.~\ref{fig4}. One can see in Figure~\ref{fig:Fig_4_DE}b, that the critical magnetic field $B_c$ shifts to lower fields when the temperature increases. Indeed, in this model, $\Delta E_Z^{\rm intr}(B_c)+\Delta E_Z^{\rm exch}(B_c)=0$. The intrinsic term, $\Delta E_Z^{\rm intr}(B)$, linearly depends on magnetic field and does not change with temperature. The exchange term, $\Delta E^{\rm exch}_Z(B)$, quickly increases in low fields and then saturates. The exchange term becomes smaller at elevated temperatures, because the average Mn$^{2+}$ spin reduces (see Fig.~\ref{fig:s5} and comments to it). Consequently, $B_c$ reduces with the increasing temperature.

Similarly, the bright exciton DCP follows the same trend. Fig.~\ref{fig:Fig_4_BE} shows calculated DCP of the bright exciton as a function of magnetic field at various temperatures in samples \#1--\#6 with parameters described above. $B_c$ is larger than for dark exciton and weakly depends on temperature. This is due to the following. First, the bright exciton intrinsic $g$-factor $g_{\rm XA}=-g_e-3g_h=+0.4$ is an order of magnitude smaller than the dark exciton $g$-factor ($g_{\rm XF}=g_e-3g_h=+3.8$). Second, the exchange term for bright exciton is several times larger than for the dark exciton: $\left(N_0\alpha f_e - N_0\beta f_h\right) / \left(-N_0\alpha f_e - N_0\beta f_h\right)=1.5-2.4$ in samples \#2--\#6. This means that $B_c$ is much larger for the bright exciton than for the dark exciton (for example, 62 and 1.5 T in sample \#5). Third, in a very strong magnetic field, $T=20$ K is not enough to depolarize the Mn$^{2+}$ ions. As a result, $B_c$ shifts to lower fields with increasing temperature, but this shift is very small (see additional panels in Figs.~\ref{fig:Fig_4_BE}d--f).

%\cl{Show how exciton energy levels depend on temperature? The dependence is not strong}

%\vspace{15mm}
\clearpage
\textbf{\large S6. Comment on anticrossing of exciton levels in a magnetic field}
%\vspace{3mm}

Calculated exciton energy levels shown in insets in Fig.~\ref{fig4} cross. This can lead to anticrossing of exciton energy levels, \latin{i. e.} a strong resonance mixing of dark $\ket{\pm2}$ and bright $\ket{\pm1}$ excitons. This effect can significantly influence the intensity and polarization of the PL.\cite{Ivchenko2018_s}

\begin{figure*}[h!]
	\begin{center}
		\includegraphics{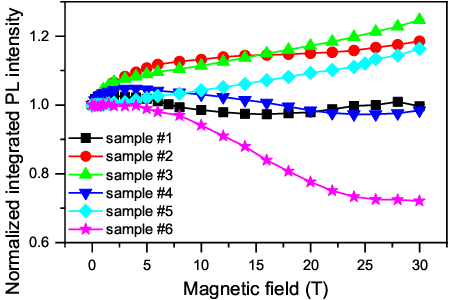}
		\caption{\label{fig:s13int}\small Normalized integrated PL intensity as a function of magnetic field at $T=4$ K in samples \#1--\#6.}   
	\end{center}
\end{figure*}

However, we observe no sharp resonances in dependences of the DCP (Figs.~\ref{fig2} and \ref{fig3}) and PL intensity (Fig.~\ref{fig:s13int}) on a magnetic field. This indicated that the effect of mixing is weak and can be neglected.

\vspace{15mm}
%\clearpage
\textbf{\large S7. Comment on spin-dependent recombination}
%\vspace{3mm}

\begin{figure*}[h!]
	\begin{center}
		\includegraphics{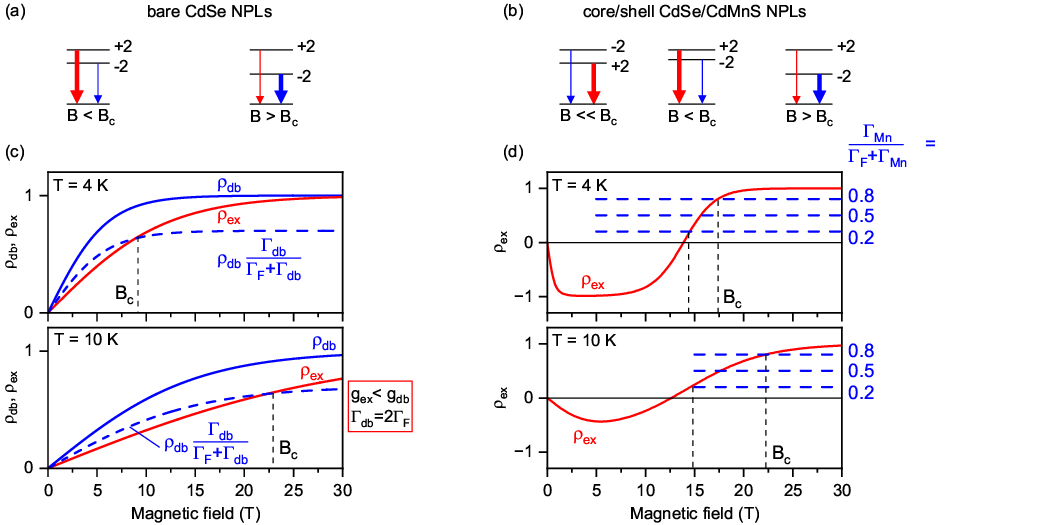}
		\caption{\label{fig:s12db}\small (a) Schematic representation of spin-dependent recombination in bare CdSe NPLs. Left: In a magnetic field $B<B_c$, the exciton with angular momentum projection $\ket{-2}$ has lower energy and higher occupation, but the spin-dependent recombination is favorable for the emission of excitons with $\ket{+2}$ angular momentum projections, and the PL is $\sigma^+$ polarized. Right: In a magnetic field $B>B_c$, the Zeeman splitting of the exciton is strong, and $\sigma^-$ polarized emission of thermally populated $\ket{-2}$ exciton level overcomes the spin-dependent recombination. (b)~Schematic representation of spin-dependent recombination in DMS core/shell NPLs. Left: In a low magnetic field $B \ll B_c$, the exciton with angular momentum projection $\ket{+2}$ has lower energy and higher occupation. Spin-dependent recombination cannot be observed. Middle: In a magnetic field $B<B_c$, the exciton with angular momentum projection $\ket{-2}$ has lower energy and higher occupation, but the spin-dependent recombination is favorable for the emission of excitons with $\ket{+2}$ angular momentum projections, and the PL is $\sigma^+$ polarized. Right: In a magnetic field $B>B_c$, the Zeeman splitting of the exciton is strong, and $\sigma^-$ polarized emission of thermally populated $\ket{-2}$ exciton level overcomes the spin-dependent recombination. (c) Explanation of critical magnetic field $B_c$ in bare CdSe NPLs. Exciton polarization $\rho_{ex}$ (red) is calculated for $g_{ex}=1$. Surface spin polarization $\rho_{db}$ (blue, solid line) is calculated for spin $g_{\rm Mn}$. $\rho_{db} \Gamma_{db}/\left(\Gamma_F+\Gamma_{db}\right)$ (blue, dashed line) is calculated for $\Gamma_{db}=2\Gamma_F$. (d) Critical magnetic field under assumption of spin-dependent recombination in DMS NPLs. $\rho_{ex}$ is calculated from exciton Zeeman splitting (see text). Average spin of Mn is assumed to be $-5/2$ in $B>5$~T at $T=4$~K and in $B>15$~T at $T=10$~K. Dashed lines correspond to $\Gamma_{db}/\left(\Gamma_F+\Gamma_{db}\right)=0.8$, $0.5$, and $0.2$. The shift of $B_c$ to higher fields with increasing temperature occurs only for very strong spin-dependent recombination.}   
	\end{center}
\end{figure*}

A non-monotonous DCP dependence on magnetic field, with the sign reversal from positive to negative, has been observed in nonmagnetic bare CdSe NPLs.\cite{Shornikova2020nn_s} The effect is caused by the spin-dependent recombination of the dark exciton assisted by flips of the surface spins. The surface spins are polarized by magnetic field, making the emission of excitons with angular momentum projection $\ket{+2}$, which is $\sigma^+$ polarized, more favorable than the emission of excitons with angular momentum projection $\ket{-2}$, which is $\sigma^-$ polarized. In low magnetic fields, this effect can overcome the intrinsic $\sigma^-$ polarized emission caused by thermal population of the Zeeman levels. In other words, while the exciton level with momentum projection $\ket{-2}$ has lower energy and higher occupation, the emission from the upper $\ket{+2}$ level is stronger due to the spin-dependent recombination \latin{via} surface spins, see Fig.~\ref{fig:s12db}a. When magnetic field increases, the Zeeman splitting grows, and the exciton polarization increases. This is favorable for the intrinsic $\sigma^-$ polarized emission. Therefore, the DCP is positive in low magnetic fields, and switches the sign to negative in high magnetic fields. The critical magnetic field $B_c$, at which it happens, can be found from the following condition: $\rho_{ex}=\rho_{db}\Gamma_{db}/\left(\Gamma_F+\Gamma_{db}\right)$, where $\rho_{ex}$ and $\rho_{db}$ are the dark exciton and surface spin polarization, respectively, and $\Gamma_{db}$ is the rate of surface spin-assisted recombination. Figure~\ref{fig:s12db}c shows an example of this behavior for bare CdSe NPLs. The top panel shows $\rho_{ex}$ (red), $\rho_{db}$ (blue, solid line), and $\rho_{db}\Gamma_{db}/\left(\Gamma_F+\Gamma_{db}\right)$ (blue, dashed line) for $T=4$ K. We assumed dark exciton $g$-factor $g_{ex}=1$ and $\Gamma_{db}=2\Gamma_F$. The critical magnetic field $B_c$ can be found as the crossing point of $\rho_{ex}$ and $\rho_{db}\Gamma_{db}/\left(\Gamma_F+\Gamma_{db}\right)$. The bottom panel shows the same for $T=10$ K. $B_c$ shifts to higher magnetic field with increasing temperature.

Below we discuss the possibility of spin-dependent recombination in DMS NPLs. 

Mn$^{2+}$ ions can act similarly to surface spins and assist the dark exciton recombination. In a magnetic field, Mn$^{2+}$ spins are polarized, and the recombination involving these spins is spin-dependent. Mn$^{2+}$ polarization depends on the Mn concentration, temperature, and magnetic field (Fig.~\ref{fig:s5}b). The spin of Mn$^{2+}$ is $5/2$. This state is degenerate in zero magnetic field, and splits in a magnetic field into six levels with spin projections from $\ket{-5/2}$ to $\ket{+5/2}$ separated by energies of $g_{\rm Mn} \mu_B B$. For small Mn concentration, low temperature, and high magnetic field, the average spin is $-5/2$, which means that all Mn spins are at the lowest $\ket{-5/2}$ level. This is favorable for recombination of dark excitons with spin projection $\ket{+2}$. In DMS NPLs in low magnetic fields, the $\ket{+2}$ exciton level has lower energy and is thermally populated. The resulting PL is $\sigma^+$ polarized (Fig.~\ref{fig:s12db}b, left), and we cannot conclude from the sign of PL, whether it is determined by thermal population of the $\ket{+2}$ level, or it is polarized due to spin-dependent recombination.

Figure~\ref{fig:s12db}d shows calculated $\rho_{ex}$ of dark excitons in sample \#2 (red). For the calculation, it is supposed that $\Delta E_{AF}=\infty$ (\textit{i. e.} only dark excitons are involved in recombination), there is recombination \textit{via} Mn$^{2+}$ spins, and other parameters are described in the main text. Exciton polarization is $\rho_{ex}=\left(n_{ex}^{-2}-n_{ex}^{+2}\right)/\left(n_{ex}^{+2}+n_{ex}^{-2}\right)$, where $n_{ex}^{+2}$ and $n_{ex}^{-2}$ are occupations of the $\ket{+2}$ and $\ket{-2}$ exciton levels, respectively. Here we choose the sign of $\rho_{ex}$ in a way, that at $\rho_{ex}>0$ the $\ket{-2}$ exciton level has lower energy. We assume Boltzmann population of the $\ket{+2}$ and $\ket{-2}$ levels, and calculate the splitting between them using eq. \eqref{eq:s_DE} (the energy levels for $T=4$ K are shown in inset in Fig.~\ref{fig:Fig_4_DE}b). In low magnetic fields, $\rho_{ex}<0$, which corresponds to the left scheme in Fig.~\ref{fig:s12db}b (the $\ket{+2}$ level has lower energy). When magnetic field increases, $\rho_{ex}$ crosses zero and reverses the sign, but the PL is $\sigma^+$ polarized due to the spin-dependent recombination \textit{via} Mn$^{2+}$ spins (Fig.~\ref{fig:s12db}b, middle). In high magnetic fields, the splitting between the lower $\ket{-2}$ and the upper $\ket{+2}$ levels increases, the lower $\ket{-2}$ level has population much higher than the upper $\ket{+2}$ level, which cannot be compensated by Mn-assisted recombination, and the PL is $\sigma^-$ polarized (Fig.~\ref{fig:s12db}b, right).

In this model, the critical magnetic field can be found from the following condition:
\begin{equation}
	\rho_{ex}(B_c)=p_{+,\rm Mn}\Gamma_{db}/\left(\Gamma_F+\Gamma_{db}\right),
\end{equation}
where $p_{+,\rm Mn}$ is the probability to find a Mn$^{2+}$ spin projection, which is suitable for recombination of $\ket{+2}$ excitons. Note that an exciton with the angular momentum projection $\ket{+2}$ can recombine by flipping Mn with spin projections $\ket{-5/2}$, $\ket{-3/2}$, $\ket{-1/2}$, $\ket{+1/2}$, and $\ket{+3/2}$, while an exciton with the angular momentum projection $\ket{-2}$ can recombine by flipping Mn with spin projections $\ket{-3/2}$, $\ket{-1/2}$, $\ket{+1/2}$, $\ket{+3/2}$, and $\ket{+5/2}$. The distribution of Mn spins depends on Mn concentration, temperature and magnetic field. We suppose that $p_{+,\rm Mn}\approx1$ in magnetic fields $B>5$ T at $T=4$ K, and $B>15$ T at $T=10$ K. This assumption slightly increases the $B_c$, but is reasonable enough to provide qualitatively correct results. Dashed lines in Fig.~\ref{fig:s12db}d correspond to $p_{+,\rm Mn}=1$, and $\Gamma_{db}/\left(\Gamma_F+\Gamma_{db}\right)=0.8$, $0.5$, and $0.2$. The magnetic field, at which $\rho_{ex}$ crosses the dashed line is $B_c$. If $\Gamma_{db}/\left(\Gamma_F+\Gamma_{db}\right)=0.2$, the shift of $B_c$ is very small. The shift increases, when  $\Gamma_{db}/\left(\Gamma_F+\Gamma_{db}\right)$ increases. The shift of about 5 T is predicted only if the spin-depend recombination is very strong, at $\Gamma_{db}/\left(\Gamma_F+\Gamma_{db}\right)=0.8$, which means that $\Gamma_{db}=4\Gamma_F$. This would mean that, in DMS NPLs the recombination is about five times faster than in nonmagnetic NPLs ($\Gamma_{db}+\Gamma_F=5\Gamma_F$ \textit{versus} $\Gamma_F$, respectively). We believe that, if present, this effect could be observed easily in experiments, for example, in lifetime measurements. However, we observe no dramatic acceleration of emission dynamics in DMS samples \#2--\#6 compared to the nonmagnetic sample \#1. We also did not observe it in our previous studies on similar NPLs.\cite{Shornikova2020acsn_s} The requirement for very strong Mn-assisted recombination is partially due to the fact that recombination of dark excitons not assisted by Mn would lead to $B_c$ shifting to lower fields with increasing temperature (see main text, and Section S5). The spin-dependent recombination should be very strong to compensate this effect.

To summarize, the spin-dependent recombination of dark excitons by flipping the Mn$^{2+}$ spins can describe the non-monotonous DCP and predicts that the critical magnetic field $B_c$ shifts to higher fields with increasing temperature. However, this explanation requires an assumption of extremely strong Mn-assisted recombination with a rate at least four times faster than the dark exciton decay rate~$\Gamma_F$. The experimental results contradict this assumption, since we do not observe dramatically different radiative recombination dynamics in nonmagnetic sample \#1 and DMS samples \#2--\#6.

\clearpage
\textbf{\large S8. Applicability of the model to other colloidal NCs}

\vspace{5mm}

\textbf{A. Nonmagnetic NCs with opposite signs of bright and dark exciton $g$-factors.}

\vspace{2mm}

A non-monotonous DCP and a shift of $B_c$ to higher magnetic fields with increasing temperature can be observed in non-magnetic NCs, if the bright and dark exciton $g$-factors have opposite signs.

\begin{figure*}[h!]
	\begin{center}
		\includegraphics{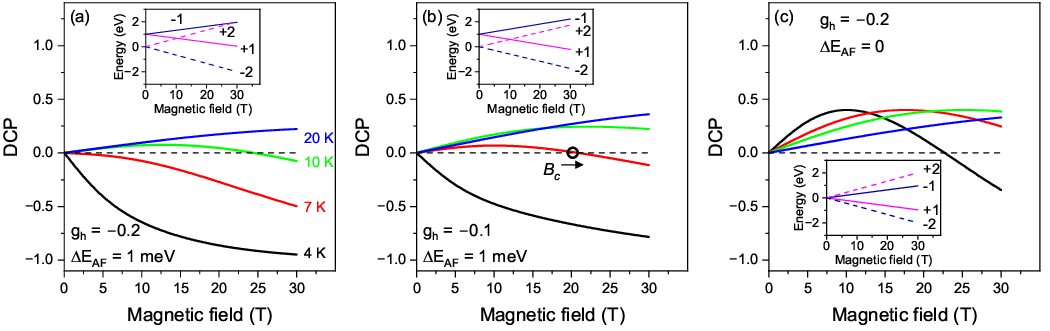}
		\caption{\label{fig:Fig_S14}\small Calculated DCP of nonmagnetic CdSe NCs at various temperatures. (a) $g_h=-0.2$, $\Delta E_{AF} = 1$ meV, (b) $g_h=-0.1$, $\Delta E_{AF} = 1$ meV, (c) $g_h=-0.2$, $\Delta E_{AF} = 0$. Other parameters are as in Fig.~\ref{fig4}. Insets show calculated exciton energies, the $\sigma^-$ emitting states are dark blue, and the $\sigma^+$ emitting states are pink.}   
	\end{center}
\end{figure*} 

Figures~\ref{fig:Fig_S14}a,b show calculated DCP for a nonmagnetic sample with $\Delta E_{AF} = 1$ meV and small negative hole $g$-factors: $g_h=-0.2$ and $g_h=-0.1$, respectively. The corresponding exciton levels are shown in insets. The DCP is negative at 4 K, because the the lowest dark exciton state is $\ket{-2}$ and the bright exciton state is not populated. With increasing temperature, the DCP in low magnetic fields becomes positive (starting from 10 K in panel (a) and 7 K in panel (b)) due to the emission of thermally populated $\ket{+2}$ and $\ket{+1}$ states. The DCP switches the sign to negative in high magnetic fields. Fig.~\ref{fig:Fig_S14}c shows DCP calculated for $g_h=-0.2$ and $\Delta E_{AF}=0$. DCP is positive already at 4 K in low magnetic fields, where emission of the $\ket{+1}$ states overcomes the $\ket{-2}$ state. DCP switches the sign in high magnetic fields, when the separation between the  $\ket{+1}$ and $\ket{-2}$ states increases. $B_c$ increases with increasing temperature in all three panels.

\vspace{5mm}

\textbf{B. Influence of bright-dark splitting.}

\vspace{2mm}

\begin{figure*}[h!]
	\begin{center}
		\includegraphics{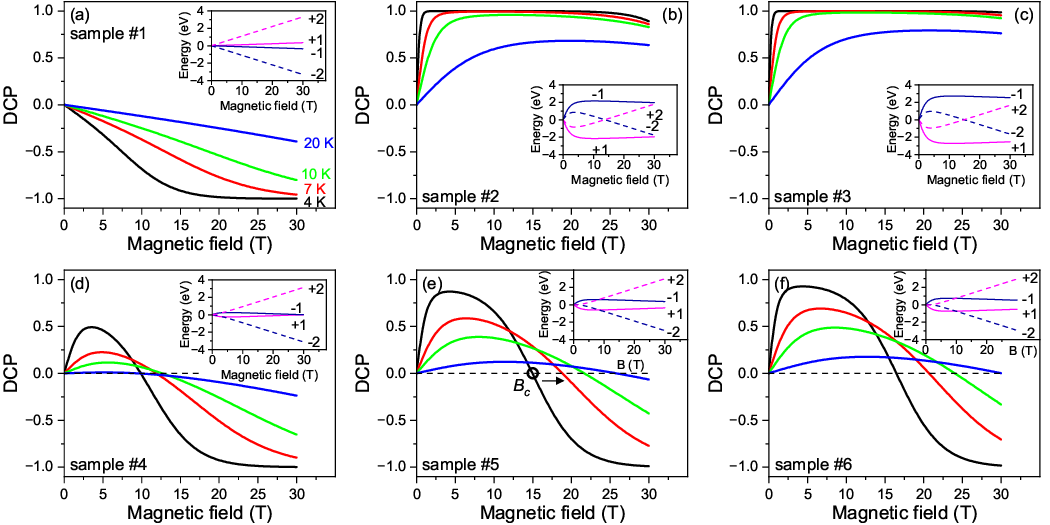}
		\caption{\label{fig:Fig_S15}\small Calculated DCP of excitons as a function of magnetic field at various temperatures for samples \#1--\#6 with $\Delta E_{AF}=0$ and other parameters as in Fig.~\ref{fig4}. Insets show calculated exciton energies at $T=4$ K, the $\sigma^-$ emitting states are dark blue, and the $\sigma^+$ emitting states are pink.}   
	\end{center}
\end{figure*}

\begin{figure*}[h!]
	\begin{center}
		\includegraphics{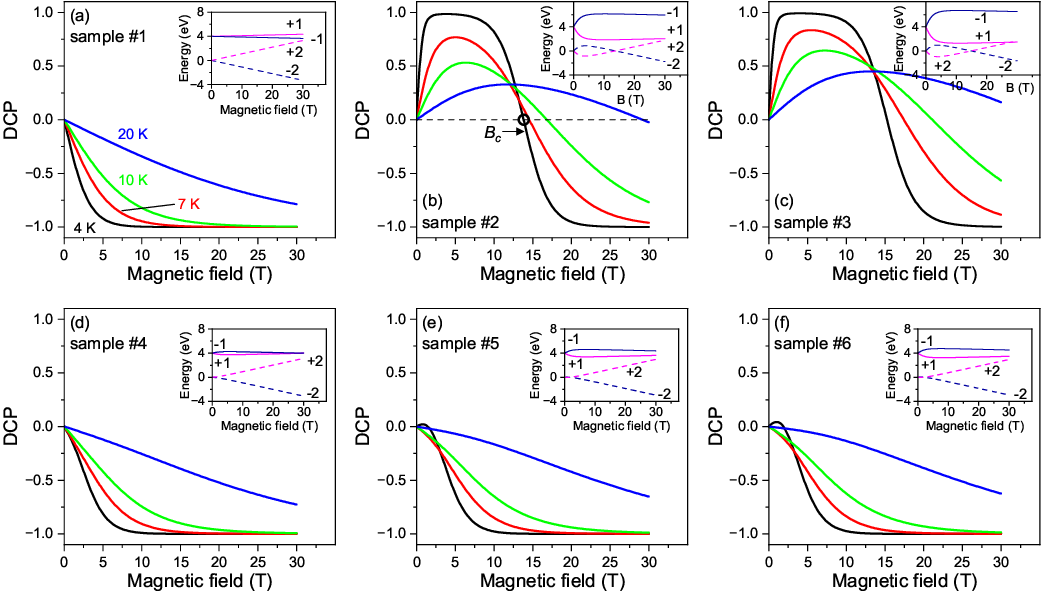}
		\caption{\label{fig:Fig_S16}\small Calculated DCP of excitons as a function of magnetic field at various temperatures for samples \#1--\#6 with $\Delta E_{AF}=4$ meV and other parameters as in Fig.~\ref{fig4}. Insets show calculated exciton energies at $T=4$ K, the $\sigma^-$ emitting states are dark blue, and the $\sigma^+$ emitting states are pink.}   
	\end{center}
\end{figure*}

\begin{figure*}[h!]
	\begin{center}
		\includegraphics{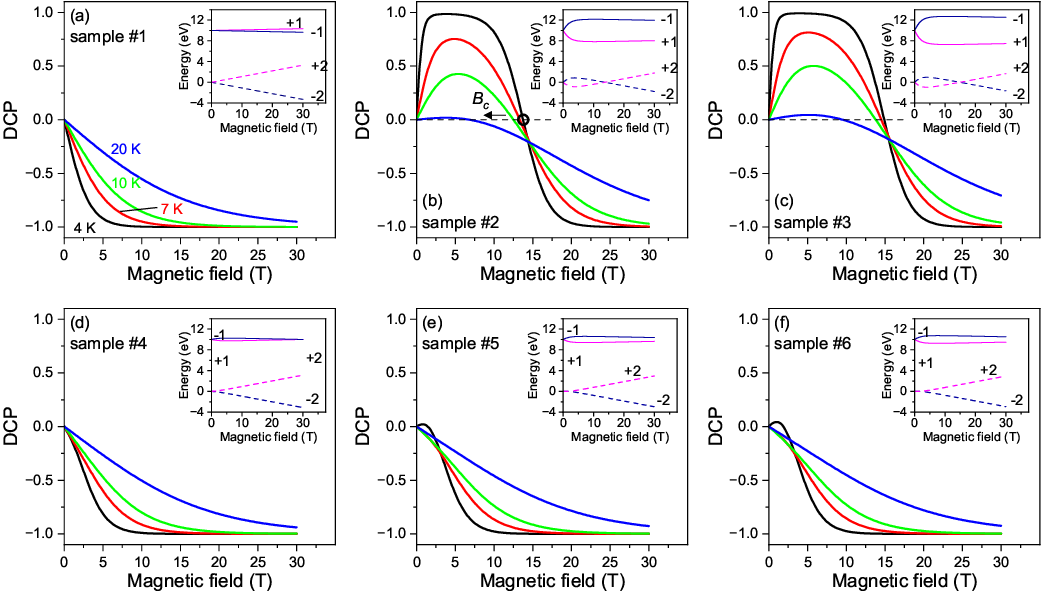}
		\caption{\label{fig:Fig_S17}\small Calculated DCP of excitons as a function of magnetic field at various temperatures for samples \#1--\#6 with $\Delta E_{AF}=10$ meV and other parameters as in Fig.~\ref{fig4}. Insets show calculated exciton energies at $T=4$ K, the $\sigma^-$ emitting states are dark blue, and the $\sigma^+$ emitting states are pink.}   
	\end{center}
\end{figure*}

A shift of $B_c$ to higher fields due to the bright-dark exciton interplay can be observed if the bright-dark splitting $\Delta E_{AF}$ is relatively small. Figures~\ref{fig:Fig_S15}, \ref{fig:Fig_S16}, and \ref{fig:Fig_S17} show DCP calculated for various $\Delta E_{AF}$ (0, 4, and 10 meV, respectively). The smaller is $\Delta E_{AF}$ the larger is $B_c$, for example, in sample \#5 it is 15 T at $\Delta E_{AF}=0$ and 2 T at $\Delta E_{AF}=4$ meV. At large $\Delta E_{AF}$ $B_c$ shifts to lower fields with increasing temperature, compare Figs.~\ref{fig:Fig_S15}b, \ref{fig:Fig_S16}b, and \ref{fig:Fig_S17}b.

\clearpage

%\clearpage
{}	
	
\end{document}